%% file: main.tex
\documentclass[12pt]{iopart}

\usepackage{graphicx}
\usepackage{color}
\usepackage{epsfig}
\usepackage{float}
\usepackage{multirow}
\expandafter\let\csname equation*\endcsname\relax
\expandafter\let\csname endequation*\endcsname\relax
\usepackage{amsfonts}
\usepackage{amssymb}
\usepackage{amsmath}
\usepackage{tabularx,url,color}
\usepackage{hyperref}
\usepackage{ulem}
\usepackage{subfigure}
\usepackage{lineno}
\usepackage{upgreek}
\usepackage{comment}
\usepackage{xcolor}
\usepackage{tablefootnote}

\input{definitions}

\begin{document}
\leftline{Dated: \today}

\title{Calibration of Advanced Virgo and Reconstruction of the detector strain $h(t)$ during the Observing Run O3}
\author{%
F~Acernese$^{1,2}$, 
M~Agathos$^{3}$, 
A~Ain$^{4}$, 
S~Albanesi$^{5}$, 
A~Allocca$^{6,2}$, 
A~Amato$^{7}$, 
T~Andrade$^{8}$, 
N~Andres$^{9}$, 
T~Andri\'c$^{10}$, 
S~Ansoldi$^{11,12}$, 
S~Antier$^{13}$
% ...
\\
{(The Virgo collaboration)}%
}%
\medskip
\address {$^{1}$Dipartimento di Farmacia, Universit\`a di Salerno, I-84084 Fisciano, Salerno, Italy }
\address {$^{2}$INFN, Sezione di Napoli, Complesso Universitario di Monte S. Angelo, I-80126 Napoli, Italy }
\address {$^{3}$Theoretisch-Physikalisches Institut, Friedrich-Schiller-Universit\"at Jena, D-07743 Jena, Germany }
\address {$^{4}$INFN, Sezione di Pisa, I-56127 Pisa, Italy }
\address {$^{5}$INFN Sezione di Torino, I-10125 Torino, Italy }
\address {$^{6}$Universit\`a di Napoli ``Federico II'', Complesso Universitario di Monte S. Angelo, I-80126 Napoli, Italy }
\address {$^{7}$Universit\'e de Lyon, Universit\'e Claude Bernard Lyon 1, CNRS, Institut Lumi\`ere Mati\`ere, F-69622 Villeurbanne, France }
\address {$^{8}$Institut de Ci\`encies del Cosmos (ICCUB), Universitat de Barcelona, C/ Mart\'i i Franqu\`es 1, Barcelona, 08028, Spain }
\address {$^{9}$Laboratoire d'Annecy de Physique des Particules (LAPP), Univ. Grenoble Alpes, Universit\'e Savoie Mont Blanc, CNRS/IN2P3, F-74941 Annecy, France }
\address {$^{10}$Gran Sasso Science Institute (GSSI), I-67100 L'Aquila, Italy }
\address {$^{11}$Dipartimento di Scienze Matematiche, Informatiche e Fisiche, Universit\`a di Udine, I-33100 Udine, Italy }
\address {$^{12}$INFN, Sezione di Trieste, I-34127 Trieste, Italy }
\address {$^{13}$Universit\'e de Paris, CNRS, Astroparticule et Cosmologie, F-75006 Paris, France }
% ...

\date{\today}

\begin{abstract}
The three Advanced Virgo and LIGO gravitational wave detectors participated to the third observing run (O3) between 1~April~2019 15:00~UTC and 27~March~2020 17:00~UTC,
leading to \dv{several gravitational wave detections per month}. 
This paper describes the Advanced Virgo detector calibration 
and the reconstruction of the detector strain $h(t)$ during~O3,
as well as the estimation of the associated uncertainties.
For the first time, the photon calibration technique as been used as reference for Virgo calibration, which allowed to cross-calibrate the strain amplitude of the Virgo and LIGO detectors.
The previous reference, so-called free swinging Michelson technique,
has still been used but as an independent cross-check. 
$h(t)$~reconstruction and noise subtraction were processed online, 
with good enough quality to prevent the need for offline reprocessing,
except for the two last weeks of September~2019.
The uncertainties for the reconstructed $h(t)$~strain, estimated in this paper \dv{in a 20-2000~Hz frequency band}, are 
frequency independent: 
5\%~in amplitude,
35~mrad in phase 
and 10~\mus\ in timing, 
with the exception of larger uncertainties around 50~Hz.

\end{abstract}

\maketitle

\tableofcontents


\input{intro}

\input{detector}


\input{calib}

\input{hrec}

\input{conclusion}


\section*{Acknowledgements}
\input{acknowledgments}

\section*{References}

\bibliographystyle{iopart-num}
\bibliography{references}

\end{document}

%% file: intro.tex
\section{Introduction}

The Advanced Virgo detector~\cite{TDR,TheVirgo:2014hva} is located near Pisa (Italy) and is looking for gravitational waves  
sources emitted by astrophysical compact sources in the frequency range 10~Hz to a few~kHz.
The O3 observing run started on $1^{\text{st}}$ April 2019 and ended on $27^{\text{th}}$ March 2020. The network of detectors was composed of the Advanced Virgo interferometer~\cite{bib:2019_SQZ_O3_PhRvL.123w1108A,bib:2020_SQZ_source_galaxies8040079}
and of the two Advanced LIGO \dv{interferometers}~\cite{0264-9381-32-7-074001,PhysRevD.102.062003,bib:ligoreco}.
\dv{Data from} the three detectors have been used together to search for gravitational wave sources~\cite{Aasi:2013wya}.
\dv{During this run, 56~candidate gravitational-wave events were identified by low latency compact binary coalescence searches using data from at least one of the three detectors~\cite{bib:GraceDB}.}
The run has been divided into two periods of about six months each, O3a \dv{($1^{\text{st}}$ April 2019 to $30^{\text{th}}$ September 2019)} and O3b \dv{($1^{\text{st}}$ November 2019 to $27^{\text{th}}$ March 2020)}.
The catalog of the detections made until the end of O3a contains \dv{the first 39~events of the run}~\cite{bib:GWTC2}.
\ \\
The Advanced Virgo optical configuration for O3 consisted in a
power-recycled interferometer with 3~kilometers long
high finesse (about 450) Fabry-Perot cavities in the arms, monolithic suspensions and mirrors thermal compensation system~\cite{TDR,TheVirgo:2014hva}.
Signal recycling~\cite{bib:SignalRecycling} was not implemented yet.
\ \\
The gravitational wave strain couples to the longitudinal length degree of freedom of the interferometer. To operate the interferometer, the relative positions of the different mirrors are precisely controlled~\cite{TDR}.
In the control bandwidth, up to a few hundred hertz, the interferometer response to gravitational wave is modified.
Above a few hundred hertz, the suspended mirrors behave as free falling masses around their position. 
The length variations induced by a passing gravitational wave induce power variations at the output of the interferometer.
\ \\
The main purpose of the Virgo calibration is to reconstruct over a wide frequency band the dimensionless amplitude $h(t)$ of the detector strain from the knowledge of both the output signal and the controls of the interferometer.
The detector strain describes the projection of the gravitational wave strain onto the Virgo interferometer.

In the long wavelength
approximation~\cite{bib:Rakhmanov_2008_shortwavelengthapprox},
the differential length of the interferometer arms, $\Delta L = L_x - L_y$, is related to the detector strain $h$ by:
\begin{eqnarray}
h=\frac{\Delta L}{L_0}\quad\quad\text{where}\quad L_0\ = \ 3\ \text{km}
\label{eq:1}
\end{eqnarray} 
For coherent search of gravitational waves with multiple detectors, 
the sign of $h(t)$ must be well defined across detectors.
For Virgo, $L_x$ and $L_y$ respectively stand for the North and the West arm lengths.
\ \\
The principle of the $h(t)$ reconstruction is the same as that described in~\cite{Accadia:2014fbz,bib:2018CQGra..35t5004A}: to remove from the dark fringe output signal the contributions
of the controls signals. This requires to calibrate the responses of the mirror actuators, the detection photodiodes readout electronics and the interferometer optical response.
\ \\
Absolute timing is also a critical parameter to estimate the direction of the gravitational wave source in the sky.
Since the typical timing accuracy of the searches is of the order of 0.1~ms~\cite{Localization:2018_LRR}, 
absolute timing precision must be of the order of 0.01~ms or less~\cite{bib:2018CQGra..35t5004A}.
\\
\ \\
The scope of this paper is to give an overview of the Advanced Virgo
calibration and $h(t)$ reconstruction during the run~O3,
and to describe the associated systematic uncertainties.
The calibration is based on the Photon Calibrator (PCal)\dv{~\cite{bib:2020_PCalO3}\cite{bib:2016_ligo_pcal}}
and is thus different from the free Michelson technique which was developed for the initial Virgo detector, used in \dv{previous observing run} O2 \lr{(August 2017)} and described in~\cite{2011CQGra..28b5005A,Accadia:2014fbz,bib:2018CQGra..35t5004A}. \dv{A similar work has been done in LIGO \cite{bib:2020_ligo_calsystematic}.}

In section~\ref{sec:detector}, we briefly give an overview of the
Virgo detector components that are relevant for calibration during O3, 
emphasizing the differences with respect to the O2 Advanced Virgo configuration.

Section~\ref{sec:calibo2o3} describes how the calibration procedures were modified since the O2 run.
In sections~\ref{sec:calibTimingSensing} and \ref{sec:calibActuators},
we describe the calibration of the photodiode readout and mirror actuation
and in section~\ref{sec:calibuncertainties} we provide the systematic uncertainties
of each calibration step. 
Section~\ref{sec:calib_michelson_compare} is dedicated to the
comparison of the current calibration results with the results obtained with the free Michelson technique.

Section~\ref{sec:hrec} shows how the $h(t)$ values have been
reconstructed using the parameters, delays and transfer functions
determined by the calibration. 
Online linear noise subtraction have been run to reduce some noise
in the $h(t)$ channel as described in section~\ref{sec:NoiseSubtraction}.
In section~\ref{sec:hrecerrors}, we also describe the various checks done on the $h(t)$ channel and their use to estimate
the systematic uncertainties for amplitude, phase and timing of $h(t)$.\\

For O3, the online version of the $h(t)$ reconstruction has been used to provide $h(t)$ in real time and no further full reprocessing has been needed,
except for the last two weeks of~O3a.
It provided $h(t)$ with a latency of about 8~s to the low-latency
gravitational wave searches that triggered alerts to our multi-messenger partners. 
This online reconstruction was based on the calibration parameters estimated
with the data taken before the start of O3.
Except when stated otherwise, the results shown in this paper pertain to the final \lr{detector} calibration \dv{and uncertainties estimation} (based on calibration data acquired during the whole run), 
and to the online \lr{reconstructed strain} $h(t)$ (computed with initial calibration based on a limited set of calibration data).

%% file: detector.tex
\section{The Advanced Virgo detector during O3}
\label{sec:detector}

Most of the detector characteristics relevant to the calibration and reconstruction described
in~\cite{2011CQGra..28b5005A,Accadia:2014fbz} for initial Virgo and in~\cite{bib:2018CQGra..35t5004A} for O2 Advanced Virgo
were still valid for Advanced Virgo during O3.
They are briefly summarized in this section, with emphasis on the relevant modifications that have been done with respect to the O2 Advanced Virgo detector's configuration.\\

\begin{figure}[tb!]
  \begin{center}
    \includegraphics[angle=0,width=0.7\linewidth]{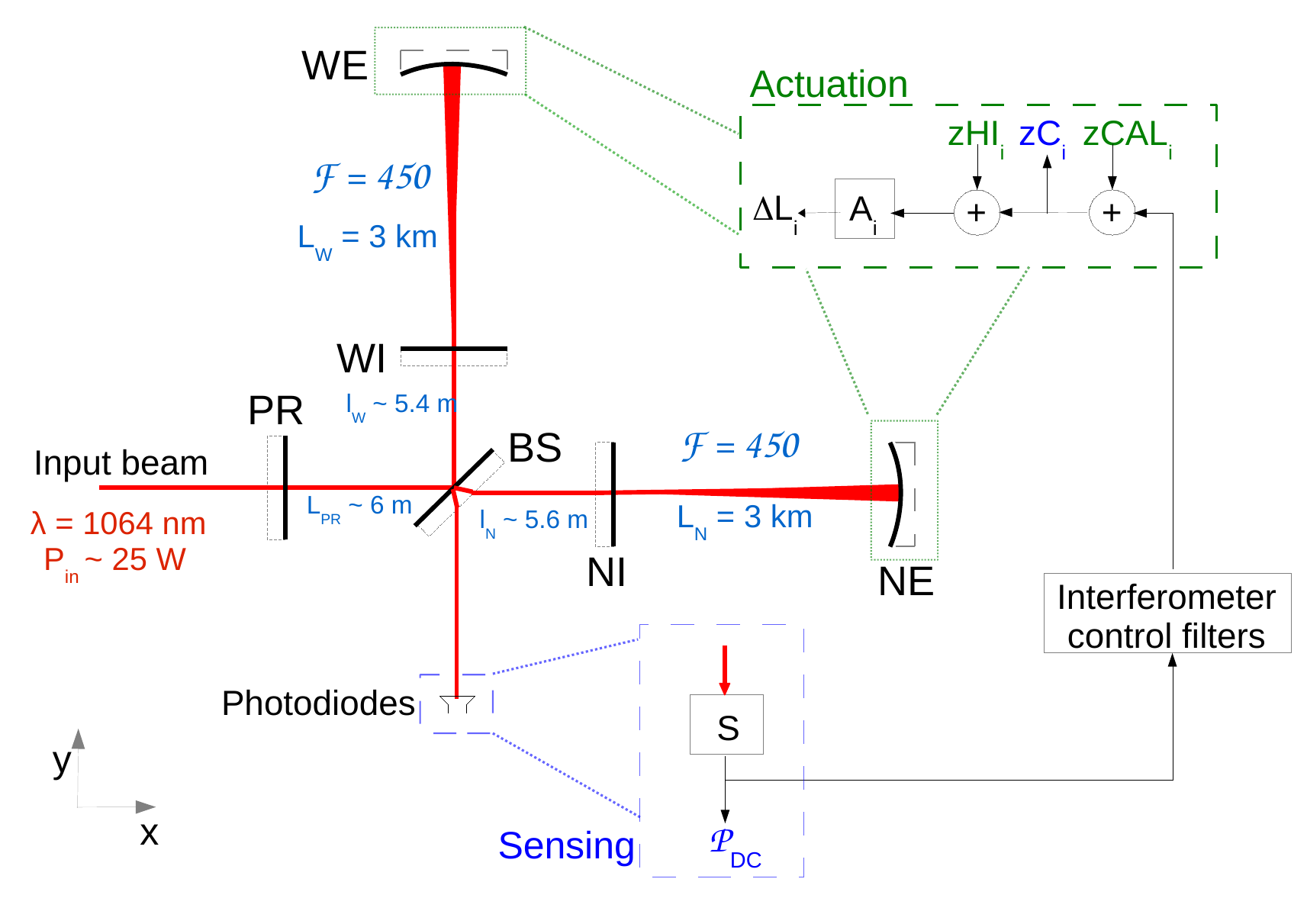}
    \caption{Configuration of the Advanced Virgo interferometer during the O3 run
	and sketch of the differential arm length control loop \msa{(so-called DARM)}. 
	\lr{The input beam is sent into the Virgo interferometer and split by the beam splitter BS towards the North (N) and West (W) arms. The arms are 3-km long Fabry-Perot cavities between the input (NI and WI) and end mirrors (NE and WE). 
	Since the interferometer is locked close to a dark fringe, the power recycling mirror (PR) is used to enhance the power on the BS mirror. 
	At the interferometer output port, photodiodes are used to measure the output power,
	with sensing response $S$, providing the channel $\mathcal{P}_{DC}$. It is used as an error signal for the differential arm length control loop. The correction signal is then used to control the NE and WE mirror positions: it is split between both mirrors $i$ whose actuation responses are $A_i$. 
	Mirror excitation can be applied via $zCAL_i$ and $zHI_i$. The channel $zC_i$ is the sum of the correction signal and $zCAL_i$, while $zHI_i$ is added later and not visible is $zC_i$.
	The main difference between both paths is about the so-called hardware injections that are described later in this paper.
	}
    }
    \label{fig:itfconfig}
  \end{center}
\end{figure}

The optical configuration of the O3 Advanced Virgo is a power-recycled interferometer with Fabry-Perot cavities as shown in figure~\ref{fig:itfconfig}.
All the mirrors of the interferometer are suspended to a chain of pendulums for seismic isolation. The last suspension stage is a monolithic silica fiber fused to the mirror on one side and attached to a marionette on the upper side (see section \ref{sec:longitudinal_actuation}). \\

The input beam is provided by a laser with a wavelength $\lambda = 1064\ \text{nm}$.
The power at the input of the interferometer during O3 was about 25~W.
The finesse of the 3-km long Fabry-Perot cavities in the arms was~450. The resulting increase of light power in the arms induced input mirrors deformation and required a Thermal Compensation System that was not used during~O2.
The readout of the interferometer main output signal is based on homodyne (or DC) detection~\cite{TheVirgo:2014hva} 
and the used photodiode signal is proportional to the interferometer differential arm length (so-called DARM). 
A small differential arm length offset is needed for the homodyne detection technique.
To reduce some noise coupling, this offset has been reduced during the Virgo commissioning period before the
start of O3, reducing the power impinging on the photodiodes.
As a consequence, the analog readout electronics has been adapted.

Frequency-independent squeezing technique~\cite{bib:2019_SQZ_O3_PhRvL.123w1108A,bib:2020_SQZ_source_galaxies8040079} was installed and used during~O3 to reduce the quantum noise limit above a few hundreds hertz.
The efficiency of the squeezing technique is limited in particular by the 
optical losses in the light detection: as a consequence, new photodiodes, with
quantum efficiency higher than 99\% were installed to readout the main output beam 
a few weeks before O3 started, along with modified analog electronics. 
The analog electronics has been again modified (implying the use of different photodiodes, even if of same type) in January~2020, a few weeks before the end of the O3~run, to reduce electronic noise. 

In order to keep a destructive interference at the interferometer output port, 
the interferometer arm length difference is controlled with a slight offset to allow for DC detection.

The other degrees of freedom are controlled by phase-modulating the laser beam at various frequencies in the MHz band 
(8~and 56~MHz among others)
and using as error signals the demodulated signals of photodiodes located at various places of the interferometer.\\
\lr{One degree of freedom of interest for calibration is the differential length of the Michelson made of the BS, NI and WI mirrors. The correction signal for this degree of freedom is applied to the BS mirror. The NI and WI mirrors were not controlled during the run~O3.}

Data from the interferometer are sampled at 10~kHz or 20~kHz and are time-stamped using the Global Positioning System (GPS).

\subsection{Mirror longitudinal actuation}
\label{sec:longitudinal_actuation}
Each Virgo mirror is suspended to a complex seismic isolation system \dv{made of a chain of seismic filters~\cite{bib:VirgoSuspension}}.
The bottom part is a double stage system with the so-called {\it marionette} as the first pendulum.
The mirror is suspended to the marionette by pairs of thin silica fibers fused to the mirror~\cite{bib:AdvancedVirgoMonolithicPayload}. 
As \dv{for the O2 run}, the reference mass is suspended to the seismic filter 
above the marionette. The resonances being below 1~Hz, the mechanical response above
10~Hz, which is to be taken into account for the calibration, has a simple $1/f^2$ behavior.\\

The positions of the marionette and mirror are adjusted with
electromagnetic actuators: permanent magnets are attached to the
marionettes and on the back of the mirrors and a set of coils
is attached to the reference mass. Electronics is used to drive the current
in every coil and to steer the suspended objects.
\lr{
On the end mirrors and beam splitter mirror, the longitudinal controls 
are distributed between the marionette (up to a few tens of Hertz) 
and the mirror (up to a few hundred Hertz).
Taking in addition the contribution of the different control signals into the detector strain,
the actuation responses of the marionette and mirror of the NE and WE mirrors need to be measured up to $\sim100\ \text{Hz}$ and up to $\sim1\ \text{kHz}$ respectively. 
The calibration of BS and PR actuators can be limited to lower frequencies.
}

\ \\
The actuation response includes the actuator itself and the
response of the suspended mirror.
The actuator is composed of a digital part, a Digital-to-Analog Converter (DAC), and the analog electronics
which converts the DAC output voltage into a current flowing through the coil.
\lr{The actuator electronics have in general different modes:
a mode with low-noise but small actuation dynamic, that is used when the interferometer is at its standard working point (so-called observing mode),
and a mode with higher actuation dynamics, but more noise, that is used during interferometer lock acquisition.
The calibration consists in calibrating the actuators in their low-noise mode, as used during observation periods.}\\
 
Between the O2 and the O3 runs, in order to reduce the thermal noise from the suspensions, 
the arm cavity payloads have been removed to replace the steel wires by the monolithic silica fibers.
The same mirrors with their electromagnetic actuators have then been put back in the interferometer.
The actuator response are then expected to be similar to the O2 response, but not identical.\\

The mirrors are sketched in figure~\ref{fig:itfconfig} and are called 
BS for the beam-splitter mirror,
WI and NI for the West and North mirrors at the input of the arm cavities,
WE and NE for the mirrors at the end of the arm cavities,
and PR for the power-recycling mirror.

\subsection{Sensing of the interferometer output power and control loops}

The main output signal of the interferometer is the power at the dark port. It is sensed using two photodiodes.
The photodiodes and their readout electronics have been changed with respect to Advanced Virgo O2 configuration, in order to adapt them to a lower impinging power and to use high quantum efficiency photodiodes.
But, as during the O2 run, the main channel  $\mathcal{P}_{DC}$, that measures the output power components from 0 to 10~kHz, is obtained from the blend of two
channels measured and digitized in two frequency bands.
The interferometer controls use demodulated channels $\mathcal{P}_{AC}$, which are extracted using digital demodulation~\cite{TheVirgo:2014hva}.

The principle of the longitudinal control loops is the same as described in~\cite{Accadia:2014fbz} or \cite{bib:2020_ISC_O3_galaxies8040085} and the sketch in figure~\ref{fig:itfconfig} 
shows the principle of the differential arm length control loop. 
Details and performances reached during~O3 are described in~\cite{bib:2020_ISC_O3_galaxies8040085}. Two feed-forward techniques setup during~O3 have nevertheless impacted the calibration, as described later, since they used the main output of the interferometer as error signal. 
The first one (so-called "Adaptive 50 Hz") is an adaptive reduction of the noise coming from the 50~Hz main power line. The second technique was setup during the commissioning break between O3a and O3b, in October 2019. It is an adaptive damping of suspension mechanical rotation resonances around 48~Hz.

%% file: calib.tex
\section{Main Virgo calibration changes since the observing run O2}
\label{sec:calibo2o3}

The parts to be calibrated for the O3 run have been the same as for the O2 run. 
Some hardware modifications and digital computation modifications of these parts were undertaken 
to improve the photodiode sensing and to reduce the thermal noise from the suspensions, 
but without significant impact on the calibration methods and outcome.

The most important update in the calibration is the change of {\it length reference}
used to calibrate the mirror actuators and hence the strain amplitude.
Until the \dv{end of the} O2 run, Virgo calibration was using the laser wavelength as reference, through the so-called
free swinging Michelson technique~\cite{bib:2018CQGra..35t5004A}. 
For O3, Virgo calibration has been done using the photon calibration technique, as in the LIGO interferometers\dv{~\cite{bib:2016_ligo_pcal}}. The technique is based on the radiation pressure of an auxiliary laser whose power must be calibrated in absolute power. 
This change of reference induced modifications in the calibration measurements and the sequence to analyse them.\\

The photon calibrator setup has been upgraded for O3. 
\dv{The PCal used during the O2 run showed large variability (about~10\%) of its power calibration~\cite{bib:2018CQGra..35t5004A}.}
Various optical modifications have permitted to reach a calibration within about 1.5\%~\cite{bib:2020_PCalO3} as summarized in more details later.

In addition, the Virgo PCal system has been cross-calibrated with the LIGO PCal system.
A systematic bias of 3.92\% in the absolute power measurements done with the Virgo powermeter has been found with respect to the LIGO power reference. It has been taken into account in the Virgo calibration procedure so that the Virgo and the LIGO length references are consistent. The final and most important outcome is to reconstruct strain $h(t)$ channels with cross-calibrated amplitudes between all detectors.\\

The free swinging Michelson technique has still been used in the O3 calibration,
but serving as an as an independent cross-check of the PCal results. 
The main limitations of this technique and the comparison with the PCal results are given in section~\ref{sec:calib_michelson_compare}.\\

Other important changes of Virgo calibration between the O2 and the O3 runs are related to the duration of the calibration periods:
while Virgo participated only to the last month of~O2, with the pre-run calibration data taken less than two weeks before,
O3 run has been the first year-long run for Virgo.
The detector was stable enough in advance
so that most of the pre-run calibration data could be taken and analysed within the two months before the start of the run.
This allowed to start the run with a reliable enough calibration.

In addition, the weekly calibration measurements spanned over a year for O3, and not only over a month as for O2. 
Hence we got more insights on calibration parameter stability.\\

\dv{A set of permanent sinewave injections which moved the mirrors at fixed frequencies and amplitudes over the whole O3 run were setup to continuously monitor the calibration and reconstruct the strain stability.
Such motions were induced with both the PCal and the electromagnetic actuators.}
An online analysis of these injections was run to generate flags and alerts in the control room in case
some calibration parameters go out of predefined bounds.\\

\dv{Another way to calibrate the Advanced Virgo interferometer}, so-called Newtonian calibration, was first tested at the end of the O2 run~\cite{bib:NCalO2_2018CQGra..35w5009E}.
The Newtonian calibrator has been improved before O3 and it could be commissioned a few times during O3.
This technique has been used as an independent cross-check of the PCal technique. 
\dv{Similar development has been started in LIGO \cite{bib:2021_ligo_ncal} and KAGRA~\cite{bib:NCal_KAGRA}.}
This technique is not discussed in this paper,
but its results, described in~\cite{bib:2020_NCalO3}, are consistent within 3\% with the PCal-based calibration, 
thus compatible with the current uncertainties.\\

To reconstruct the detector strain $h(t)$, we need to calibrate the sensing part (photodiode readout)
and the electromagnetic actuators of the controlled mirrors and marionettes.
The following sections describe the methods and results of these calibrations. 

\section{Calibration of Advanced Virgo timing and sensing}
\label{sec:calibTimingSensing}

A first part to be calibrated is the photodiode sensing: 
how the output DC power of the dark fringe beam is converted by the detection photodiodes into the main output signal~$\mathcal{P}_{DC}$. 
A model is associated that describes the transfer function of the sensing part and
the time delay introduced by the sensing.

\dv{As for the calibration of the O2~run}, the sensing model assumes a flat modulus response in the detection band, 
but with some digital anti-alias filters, dominated by a filter around 8~kHz.
Modifications of the digital processing between the O2 and the O3 runs have been accounted for.
The overall readout delay of the photodiode data acquisition has been measured by flashing \dv{a 1~PPS signal~\footnote{Pulse Per Second signal is a synchronization signal delivered by the GPS receiver}}
using a LED connected to a GPS receiver in front of the photodiodes, as during~O2.
The measured delay confirmed the expected value, with systematic uncertainties lower than~3~\mus~\cite{bib:2019_vnote_calibTiming}. 


\section{Calibration of the electromagnetic actuators using the photon calibrators}
\label{sec:calibActuators}
The calibration of Advanced Virgo mirror and marionette electromagnetic actuators consists in measuring their transfer function (modulus in m/V) that converts a known digital signal (in V) into a mirror displacement (in m) at an absolute GPS time. 

For the third observing run O3, the photon calibrators~\cite{bib:2020_PCalO3} were used for the first time as reference for the Virgo detector calibration.
\lr{To calibrate the actuators of the different mirrors and marionettes,
this reference is {\it transferred} in series from a reference to another one in different steps. 
The different measurements associated to these transfers have been taken on a weekly basis during~O3 in order to monitor the stability of the actuator calibration. \\

All the weekly measurements of a given type are combined together to assess the stability of the calibration data over a given calibration period.
As shown later, the time variations were found to be small. As a consequence,
the actuator models have been estimated using the average of all the calibration period
and the time variations have been included in the uncertainty associated to the actuator models.
The actuator model is an ad-hoc fit of the measured averaged transfer function.\\

Initial measurements were performed before the start of~O3, in March~2019, in order to derive the actuator models used online at the start of the O3a period, in the $h(t)$ reconstruction processing in particular.
Then, weekly calibration measurements have been used to monitor the stability of the calibration and to improve the statistical uncertainties of the measurements.
In October 2019, all the weekly measurements taken during O3a (April to September 2019) 
have been analysed to extract the updated actuator models that have been used online during O3b (November 2019 to March 2020). 
After O3, all the weekly measurements taken during O3 have been analysed together to extract updated actuator models and uncertainties. These models are used later in this paper to estimate uncertainties on the online strain $h(t)$ channel.
Note that these last models were not used online nor for offline reprocessing of the $h(t)$ strain signal.\\

In this section, we first explain the principle of the calibration transfer 
and the different kinds of measurements needed to transfer the photon calibration reference to all the mirror and marionette actuators.
Then, we give the results of the different calibration steps. 
As the control loop of the differential arm length directly actuates the position of the NE and WE mirrors and marionettes, we describe in details their actuator calibration. We also compare the final O3 actuator models with the initial and intermediate models, extracted with a limited set of calibration data, that have been used for online processing during the O3a and O3b periods respectively.
The calibration of the other actuators is more rapidly described:
the methods are similar and the main output given in this paper are 
the final uncertainties.
Finally, section~\ref{sec:calibuncertainties} summarizes the uncertainties of the detector calibration based on the photon calibration technique.
In section~\ref{sec:calib_michelson_compare}, we highlight some improvements of the free swinging Michelson calibration technique
as well as some of its limitations. 
Then we validate the PCal-based calibration by comparing the independent outputs of both techniques.
}

\subsection{A series of calibration transfers}\label{sec:calibStrategy}

The actuator calibration is made as a series of calibration transfers between different systems,
comparing a system of reference to another one that needs to be calibrated.
It is based on the comparison of the detector's response $R$ to two different excitation paths, from the signals $Exc_{ref}$ and $Exc_{new}$ respectively applied to a calibrated actuator of reference $A_{ref}$ and to the actuator to be calibrated $A_{new}$.
The output signal $S_{out}$ of the detector is written in the frequency domain as:
\begin{equation}  \label{eq:transfer_S}
S_{out} = Exc \times A \times R    
\end{equation}
From two dataset with the different excitation paths, two transfer functions are measured:
\begin{equation}  \label{eq:transfer_Sref}
    \frac{S_{out}}{Exc_{ref}} = A_{ref} \times R 
\end{equation}
\begin{equation}  \label{eq:transfer_Snew}
    \frac{S_{out}}{Exc_{new}} = A_{new} \times R
\end{equation}
They are then combined to extract the actuator response to be calibrated:
\begin{equation}  \label{eqn:transfer_tfratio}
    A_{new} = \frac{S_{out}}{Exc_{new}} \times \Big( \frac{S_{out}}{Exc_{ref}} \Big)^{-1} \times A_{ref}
\end{equation}


The first step of the Virgo detector calibration consists in transferring the NE and WE PCal actuators calibration to the NE and WE mirror and marionette electromagnetic actuators,
with the interferometer locked at its standard working point.
Then, different transfers are done in series between the different actuators using specific configurations
of the interferometer.
The NI and WI mirror actuators are calibrated with respect to the NE and WE mirror actuators,
with the interferometer locked at its working point but with the NI and WI actuators enabled.
The PR and BS mirror actuators are then calibrated with respect to the WI mirror actuator,
locking the cavity made of the PR-BS-WI mirrors.
Finally, the BS marionette actuators are calibrated with respect to the BS mirror actuator,
with the interferometer locked at its working point.\\
The following sections describe these different steps and associated measurements.

\subsection{Calibration of the NE and WE mirror actuators}
\label{sec:calibnewe}

\subsubsection{\bf Power calibration of the photon calibrators}
\label{sec:PCalCalibration}

The reference  used for the PCal calibration is the power of the auxiliary laser reflected on the end mirrors $P_i^{ref}$ which is calibrated by a power-meter called the Virgo Integrating Sphere, itself calibrated with respect to another power-meter called LIGO Gold Standard. The LIGO Gold Standard itself is calibrated by the National Institute of Standards and Technology (NIST) so that the beam measurements are all given in absolute power~\cite{bib:LIGO_PCalO3}.\\ 

As a preliminary step, the PCal actuators for WE and NE must be calibrated.
Their calibration during~O3 is described in the paper~\cite{bib:2020_PCalO3}.
It gives the following transfer function~[m/W]:
\begin{equation}
    A_{pcal,i}(f) = \frac{2\cos(\theta)}{\text{c}} \times H(f) \times \Big( S_{pcal}(f) \Big)^{-1}
\end{equation}
with $i=\{NE,WE\}$, $\theta$ the angle of incidence of the PCal laser beam on the end mirror,
$\text{c}$ the light speed, $H(f)$ the mechanical response of the end test mass to an excitation force
and $S_{pcal}$ the sensing function of the PCal photodiode. 
The PCal laser power is modulated to generate an excitation to the mirror, 
and the reflected power $P_i^{ref}$ is recorded on a calibrated photodiode.
The displacement of the end mirror induced by the PCal actuator 
is estimated multiplying the photodiode signal digitized at 20~kHz $P_i^{ref}$ and the PCal actuator response $A_{pcal,i}(f)$. \\

As stated in ~\cite{bib:2020_PCalO3}, the NE and WE photon calibrators have first been calibrated in March 2019, before the start of~O3.\\
The WE PCal calibration was found to be stable during O3a, with systematic uncertainties 
on the WE mirror induced motion estimated to $1.35\%$.
The WE laser stopped working at the end of August~2019 and it was replaced during the run break in October~2019.
The photodiode was recalibrated before the start of O3b. This re-calibration
induced a small offset compared to the O3a power calibration, consistent with
the systematic uncertainty coming from the power measurement with the Virgo Integrating Sphere
already taken into account.
However, the WE PCal power calibration was found to drift during O3b, in correlation with the environment humidity. This drift was seen after the end of O3b and hence was not corrected in the PCal calibration nor in the online $h(t)$ reconstruction. As a consequence, it has been included as additional uncertainties on the WE PCal power calibration: systematic uncertainties on the WE mirror induced motion were estimated to $1.73\%$ during O3b.\\
The NE PCal already had shown some calibration variations with humidity during O3a.
Systematic uncertainties on the NE mirror induced motion were estimated to $1.35\%$ during O3a.
The NE laser also failed, in October~2019, and was replaced in January~2020 to be used again for O3b calibration. No offset in the NE PCal power calibration was found and the correlation with humidity was similar as during O3a. As a consequence, the systematic uncertainties on the NE mirror induced motion estimated to $1.39\%$ during O3b.\\
\lr{During the two periods when a single PCal was available, the other being faulty,
the Virgo calibration could still be monitored with the other PCal
and with the free swinging Michelson calibration technique.}

\lr{The timing of the PCal system is also calibrated so that the reconstructed displacement of the end mirror induced by the PCal is known as a function of the GPS time: the delays from the timing distribution in Virgo and from the photodiode readout electronics must be accounted for.}
The timing uncertainty on the PCal system has been estimated to 3~\mus\ during 03a~\cite{bib:2020_PCalO3}.\\
On 17 December 2019, the PCal digital anti-imaging filter has been slightly modified, 
changing \lr{the delay of} the PCal readout by 2.8~\mus. 
This modification has not been taken into account in the analysis of the calibration data
but has been taken into account as an additional uncertainty on the PCal timing during O3b
when estimating the timing error on the detector strain $h(t)$ (see section~\ref{sec:hrecerrors_conclusions}).

Once the PCal actuator has been calibrated, it is used as reference to calibrate the electromagnetic actuators of the end mirrors (NE and WE).

\subsubsection{\bf Calibration transfer from  PCals to end mirror actuators}

To derive the models of the actuators of the end mirrors $A_{mir,i}^{Sc}$ , $i = \{$WE,NE$\}$, the interferometer is set in its standard working point \lr{(so called {\it observing mode})}. 
\lr{
The electromagnetic actuators are thus in the low-noise mode and the correction signals sent to the actuator have nominal properties,
inducing the actuator temperature to also be at its nominal level.
}
The measurements of the end actuators response does not depend on different calibration transfers as it was the case with the free swinging Michelson technique.\\

We measure the PCal to end mirror transfer functions using known photon calibrator multiplets of sine wave excitations $x_{pcal,i}(f_k)$ which are applied to the end mirrors while the error signal of the interferometer $\mathcal{P}_{DC,pcal,i}$ is measured. The effect of these excitations on the error signal are then compared to excitations at the same frequencies $x_{mir,i}(f_k)$ sent with the electromagnetic actuators of the end mirrors.\\

\lr{We can write the effect of the sine waves on the error signal in the frequency domain for the different datasets:
\begin{equation}
\mathcal{P}_{DC,mir,i} = x_{mir,i} \cdot R^{ITF}_{mir,i} = zCAL_{mir,i} \cdot TF_{mir}^{CalToSc} \cdot A_{mir,i} \cdot R^{ITF}_{mir,i} 
\end{equation}
\begin{equation}
\mathcal{P}_{DC,pcal,i} = x_{pcal,i} \cdot R^{ITF}_{mir,i}  = P^{ref}_{i} \cdot A_{pcal,i} \cdot R^{ITF}_{mir,i}
\end{equation}
where $zCAL_{mir,i}$ are the generated digital signals sent to the electromagnetic mirrors actuators as external perturbations, $TF_{mir}^{CalToSc}$ is the digital transfer function from the generated calibration signals $zCAL_{mir,i}$ to the correction signals $zC_{mir,i}$, 
$A_{mir,i}$ are the electromagnetic actuator responses from $zC_{mir,i}$ to the mirror motion $\Delta L$ at a GPS time ; 
$A_{pcal,i}$ are the actuator responses from the PCal laser power read on a photodiode ($P^{ref}_{i}$) to the mirror motion $\Delta L$ at a GPS time ;
and $R^{ITF}_{mir,i}$ is the closed-loop response of the interferometer to a motion of an end mirror.\\

Thus the responses of the actuators of the end mirrors ($i$) are computed as\footnote{The brackets $\left[\cdot\right]$ point out the transfer functions directly measured from times series stored in the data.}:
\begin{equation}
A_{mir,i}(f) = \left[\frac{\mathcal{P}_{DC,mir,i}}{zCAL_{mir,i}}\right] \cdot \left[\frac{P^{ref}_{i}}{\mathcal{P}_{DC,pcal,i}}\right] \cdot A_{pcal,i}(f) \cdot \frac{1}{TF_{mir}^{CalToSc}}
\label{eqn:PCaltoEndnew}
\end{equation}
}
As shown in next paragraph, those actuators responses were measured every week during the run~O3 to check their stability over the time and to get more precise measurements combining all the datasets.

\subsubsection{\bf Stability of the NE weekly measurements during O3}
\label{sec:TimeStabilitymirNE}

\begin{figure}[!ht]
    \centering
	\includegraphics[trim={0 0cm 0 0cm},clip,width=0.8\linewidth]{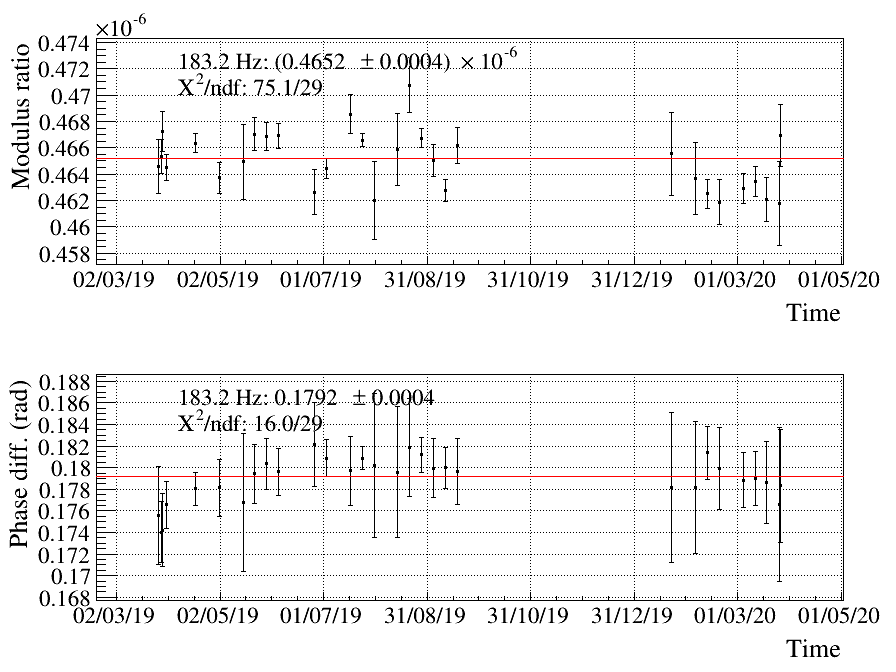} 
    \caption{Evolution of the NE transfer function ratio used to transfer PCal to electromagnetic actuator calibration as a function of time at $183.2~$Hz during O3
    (see equation~\ref{eqn:PCaltoEndnew}).
    The NE Pcal laser could not be used from  October 2019 to mid-January 2020. 
    The laser was replaced and the PCal photodiodes re-calibrated mid-January 2020.
    }
    \label{fig:mirNE_pcalstab}
\end{figure}

Figure~\ref{fig:mirNE_pcalstab} is an example of the time variation of the modulus and phase of the NE mirror actuation response at $183.2~$Hz (one of the injected calibration lines) measured every week. 
The mean value of the measurement is shown as the red line on the figure,
and its associated standard deviation and $\chi^2/ndf$ \lr{(chi squared divided by number of degrees of freedom)} are given.\\

For each $\chi^2/ndf$, one can attach a p-value which is an indicator for rejecting or not the null hypothesis. The p-value threshold above which we cannot reject the null hypothesis has been chosen to a conventional value of $0.05$.  
This means that if the p-value related to the computed $\chi^2/ndf$ is above $0.05$, the data points are consistent with a constant fit in our case. The error bars on the plots are only statistical uncertainties coming from the measurements and they are sometimes too small to have a p-value$~> 0.05$ because one needs to take into account some systematic uncertainty that shows up as time variations. We have thus implemented an iterative algorithm that adds systematic uncertainties on the statistical errors and computes the new $\chi^2/ndf$ which thus gives a new p-value until the p-value is higher than $0.05$.\\

\begin{figure}[!ht]
    \begin{minipage}[t]{0.45\textwidth}
	   \hspace{-0.5cm}    \includegraphics[width=0.99\linewidth]{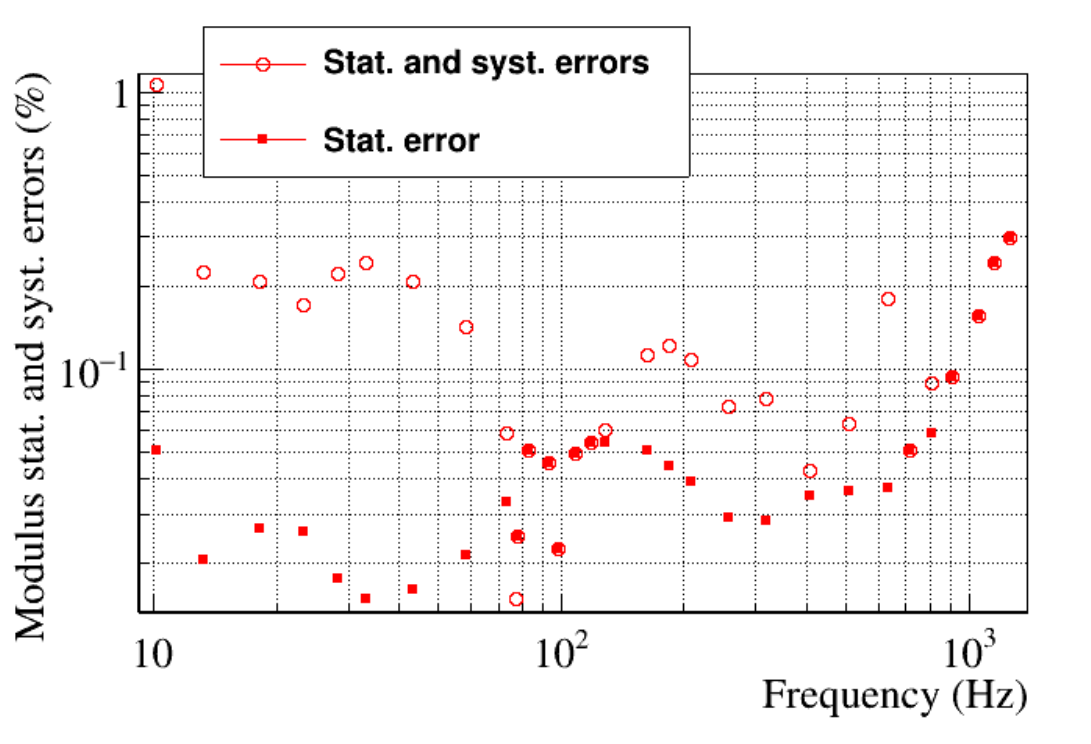} 
    \end{minipage}
    \begin{minipage}[t]{0.45\textwidth}
        \centering
	    \includegraphics[width=0.99\linewidth]{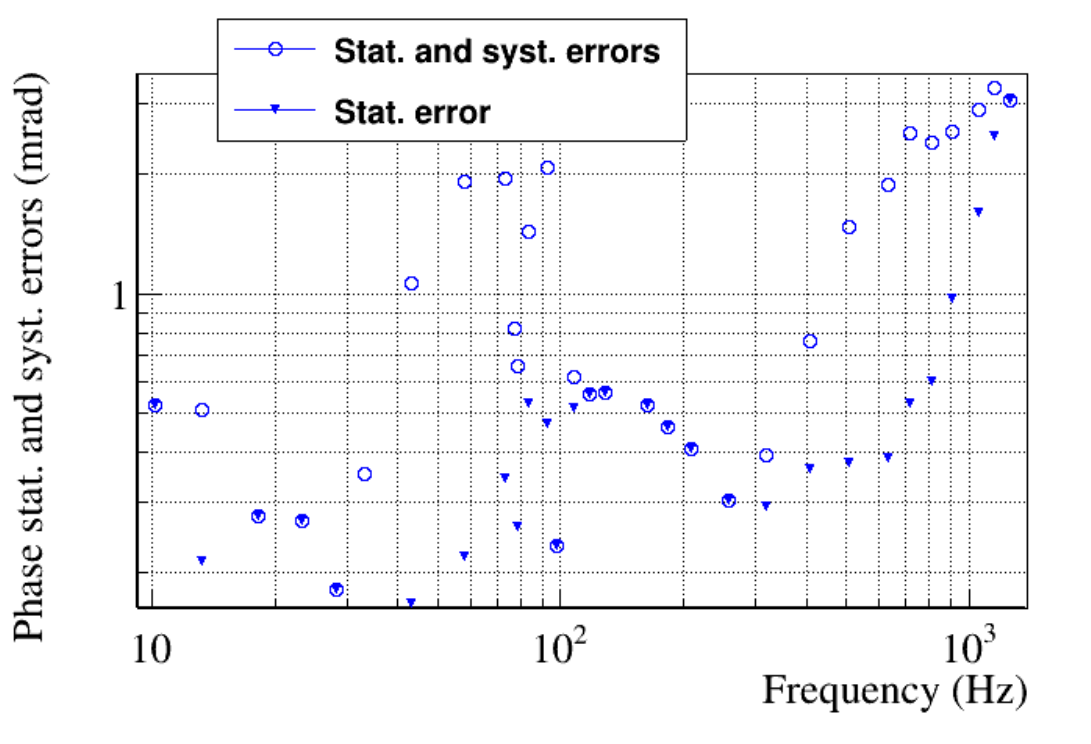}
    \end{minipage}
    \caption{Statistical and systematic errors estimation on the calibration transfer from NE PCal to NE mirror actuator. The errors on the amplitude are expressed in [$\%$] and the errors on the phase are expressed in [mrad]. Left: Statistical errors on the amplitude (red filled squares) of the calibration transfer as a function of frequency and statistical plus systematic errors (red empty circles) estimation to have a p-value$~>0.05$. Right: Statistical errors on the phase (blue filled triangles) of the calibration transfer as a function of frequency and statistical plus systematic errors (blue empty circles) estimation to have a p-value$~>0.05$.}
    \label{fig:mirNE_syst}
\end{figure}

Figure~\ref{fig:mirNE_syst} shows, as a function of frequency, the statistical errors and the statistical plus the systematic errors (linear sum) on the amplitude and phase of the calibration transfer from the photon calibrators to NE mirror actuator. Above 20~Hz, systematic errors smaller than $0.3\%$ are added to the statistical errors in amplitude on the whole frequency band. Regarding the phase, systematic errors smaller than $3~$mrad have also been added to the statistical errors so that the actuator response is compatible with a stable behavior. It is noticeable that the statistical error bars increase with the frequency, as expected since the induced mirror motion decreases while the detector sensitivity gets worst. As a consequence, possible small systematic errors are no longer visible at high frequency.\\

\lr{The conclusion of this analysis is that the NE mirror actuator responses was varying in time by less than $0.3\%$ in amplitude and less than $3~$mrad in phase.
This is fully measured in the range 20~Hz to about 1~kHz. We do not expect larger time variations at higher frequencies where the weekly data are anyway compatible within their larger statistical uncertainties.
These variations are small and their behavior between the weekly data cannot be interpolated. 
We have thus chosen to average the weekly data taken at a given frequency to get the mean modulus and phase of the actuator response for the full O3 run, and to include the estimated time variations as systematic uncertainties on the actuator response.}

\subsubsection{\bf Average of the NE weekly data and actuator model fit}

\begin{figure}[!ht]
    \centering
	\includegraphics[trim={0 0cm 0 0cm},clip,width=0.8\linewidth]{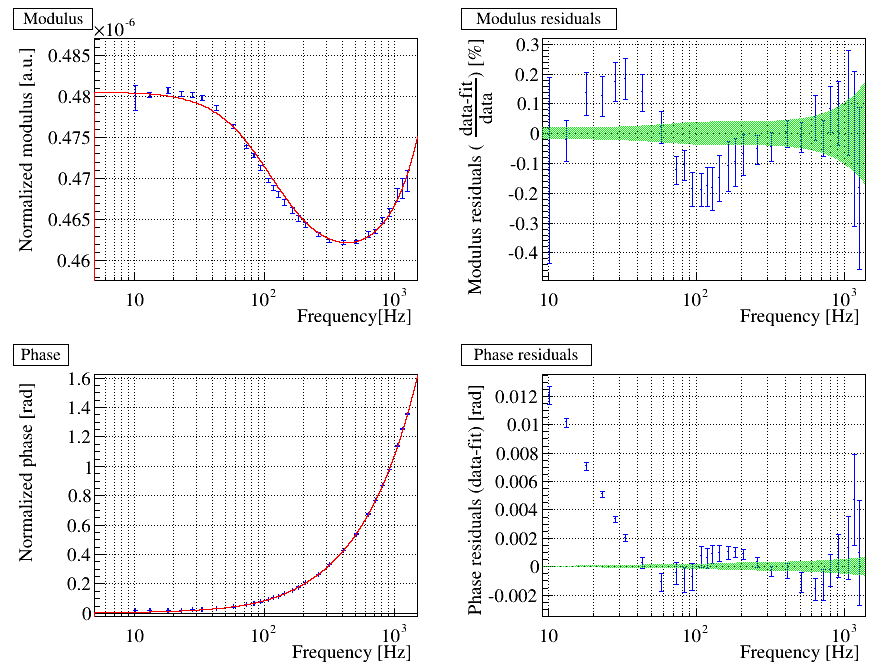} 
    \caption{NE mirror actuator response normalized by the simple pendulum mechanical model. On the left, the measurements are represented in blue and the fit is drawn in red. Modulus is given in~m/V and phase in~rad. The modulus and the phase are fitted simultaneously. 
    On the right, the residuals are shown. 
    The green area represents the uncertainties given back by the fitted model,
    taking into account parameter covariance.
    \label{fig:mirNE_fullmodel}
    }
\end{figure}

The final actuator model results from the average of all the weekly measurements done during~O3, and includes the pre-O3 measurements from March~2019. Figure~\ref{fig:mirNE_fullmodel} shows the final NE end mirror actuator response and its fit, \lr{normalized by a pole at 0.6~Hz with a quality factor of~1000.
This normalisation model approximately describes the mechanical attenuation of the suspended mirror, mainly going as $f^{-2}$ above 10~Hz. This normalisation permits to see possible small deviations of this $f^{-2}$ behavior coming from the electronics. This attenuation model does not perfectly correspond to reality, but any deviation from it would be compensated by the fit.} 

The model parameters and fit uncertainties are summarized in the last column of table~\ref{tab:mirNE_models}.
The fit residuals are within 0.2\% in modulus. In phase, some unexpected behavior is seen
and not included in the fit below 40~Hz, with the phase deviating from~0. As a consequence, these residuals  are lower than 2~mrad above 30~Hz, but reach 6~mrad at 20~Hz.
\lr{The residual deviations from the fit are taken into account as systematic uncertainties 
on the actuator model, as summarized in section~\ref{sec:calibuncertainties}.}
\\

As a rough cross-check, the actuator model has been compared to the nominal actuator response,
estimated from the actuator electronics schematics and coil-magnet geometry:
the gain is consistent within 15\%. This is not unexpected given the uncertainties on the electronics components of the actuator (resistance values, inductance values), the coil-magnet coupling factors and the balancing of the four coils.\\

\lr{The results shown in this section about the NE mirror actuator have been computed using all the O3 calibration data.
In the next section, they are compared with two intermediate calibration models 
estimated with different subset of the data.
}

\begin{figure}[!ht]
    \centering
	\includegraphics[trim={0 0cm 0 0cm},clip,width=0.8\linewidth]{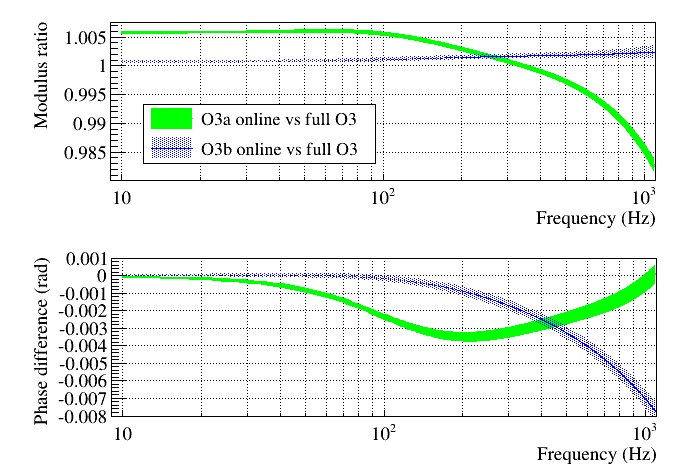}
    \caption{Comparison (ratio) of NE mirror actuator response models used online during O3 to the model extracted using the full O3 dataset (green filled area: O3a online model, blue dotted area: O3b online model).
    }
    \label{fig:mirNE_modelcomparison}
\end{figure}

\subsubsection{\bf Comparison of NE mirror final O3 model with the online models.}

\lr{Initial and intermediate} actuator models were extracted in March~2019 and in October~2019 with the dataset available at that time. They have then been used online during O3a and O3b respectively, in particular in the $h(t)$ online reconstruction processing.
The model parameters are summarized in table~\ref{tab:mirNE_models}.
Figure~\ref{fig:mirNE_modelcomparison} shows a comparison of the NE mirror actuator models used online during O3a and during O3b with the model extracted using all the O3 calibration dataset as shown in previous paragraphs. 
More dataset and better measurements around 1~kHz showed the need to add a high-frequency zero, which was not included in the O3a online model. The O3b model is in agreement with the full model within better than 0.3\% and 1~\mus\ (8~mrad at 1~kHz). Due to the absence of a high-frequency zero in the O3a online model, differences are larger.
In particular, for the modulus, the difference is up to 2\% at 1~kHz, but the agreement with the full model is still within 0.5\% up to 600~Hz.\\

\begin{table}[!ht]
\centering
\begin{tabular}{|c|c|c|c|}
\hline 
Model & O3a online &  O3b online & preO3+O3a+O3b \\ 
\hline
Fit     &  &  &  \\
Gain ($\upmu$m$/$V) & $0.4833$ & $0.4809$ & $0.4806\pm0.0001$ \\  
Pole $f_{p}$ (Hz)   & $123.5$    & $112$    &  $112.5\pm1.2$ \\ 
Zero $f_{z}$ (Hz)   & $129.8$    & $117$    &  $117.6\pm1.2$ \\ 
Zero $f_{z}$ (Hz)   & -        & $5636$     &  $5736 \pm 162$ \\
\hline
Delay ($\upmu$s)   & $-172.2$  & $-143.0$   & $-144.6\pm0.7$\\
\hline
Pendulum &  &  &     \\
$f_{0}$ (Hz) & $0.6$  & $0.6$  & $0.6$ \\
$Q_{0}$      & $1000$ & $1000$ & $1000$ \\
\hline
\end{tabular}
\caption{\label{tab:mirNE_models} NE mirror actuators models during~O3. 
The models "O3a online" were derived in March 2019 and used online during O3a ;
the models "O3b online" were derived in October 2019 using all the O3a data and were used online during O3b ;
the models "preO3+O3a+O3b" (full model) were derived after the end of O3. Their validity range is from $10~$Hz to $1500~$Hz.
}
\end{table}

\subsubsection{\bf Stability of the WE weekly measurements during O3}

The stability of the NE mirror measurements was shown in the previous sections.
We show here that the WE mirror measurements were less stable,
but that the main origin of the variations has been understood.\\

\begin{figure}[!ht]
    \centering
	\includegraphics[trim={0 0cm 0 0cm},clip,width=0.8\linewidth]{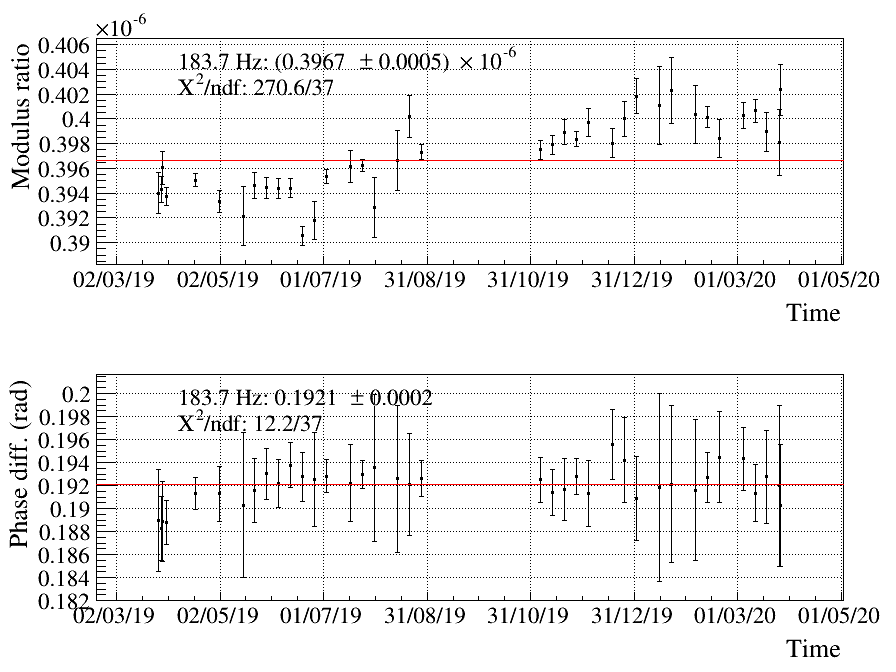} 
    \caption{Evolution of the WE transfer function ratio used to transfer PCal to electromagnetic actuator calibration as function of time at $183.2~$Hz during O3
    (see equation~\ref{eqn:PCaltoEndnew}).
    \lr{The WE Pcal laser could not be used in September 2019. The laser was replaced and the PCal photodiodes re-calibrated end of October 2019.}
    }
    \label{fig:mirWE_pcalstab}
\end{figure}

Figure~\ref{fig:mirWE_pcalstab} is an example of the time variation of the modulus and phase at $183.2~$Hz of WE mirror actuator. Some large variations are seen, in particular between O3a and O3b. As summarized in section~\ref{sec:PCalCalibration}, the laser of the PCal used as reference to calibrate the WE mirror failed at the end of August~2019 and was replaced and re-calibrated in October~2019. In addition, the PCal photodiode signal started to show up some power calibration variation with environment humidity. This explains the large variation seen in the WE mirror calibration, and in particular the differences between O3a and O3b periods. 

\begin{figure}[!ht]
    \begin{minipage}[t]{0.49\textwidth}
	   \includegraphics[width=0.99\linewidth]{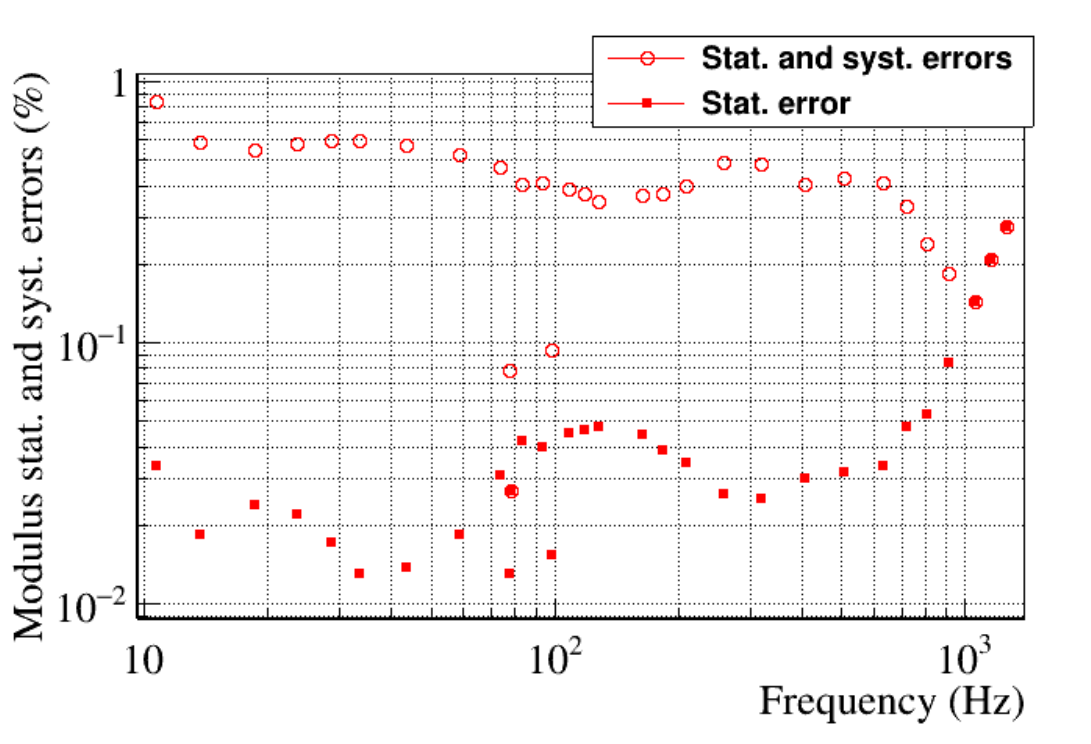} 
    \end{minipage}
    \begin{minipage}[t]{0.49\textwidth}
        \centering
	    \includegraphics[width=0.99\linewidth]{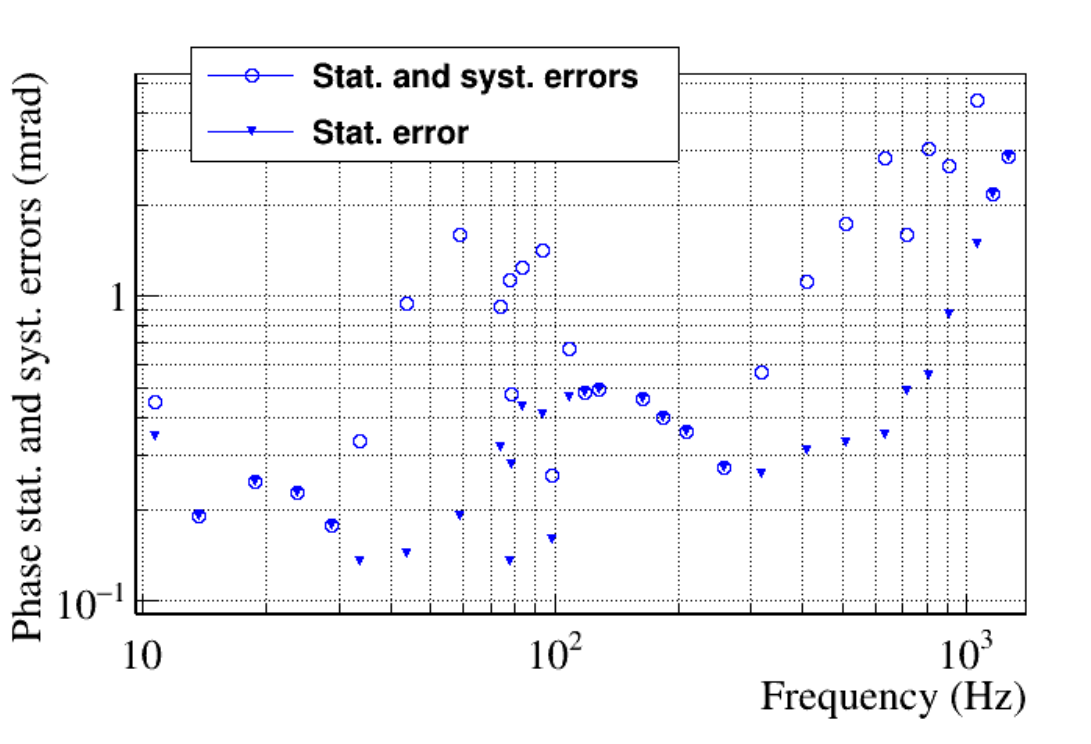}
    \end{minipage}
    \caption{Statistical and systematic errors estimation on the calibration transfer from WE PCal to WE mirror actuator. The errors on the amplitude are expressed in [$\%$] and the errors on the phase are expressed in [mrad]. Left: Statistical errors on the amplitude (red filled squares) of the calibration transfer as a function of frequency and statistical plus systematic errors (red empty circles) estimation in order to have a p-value$~>0.05$. Right: Statistical errors on the phase (blue filled triangles) of the calibration transfer as a function of frequency and statistical plus systematic errors (blue empty circles) estimation to have a p-value$~>0.05$.}
    \label{fig:mirWE_syst}
\end{figure}

We processed the same analysis as for NE on the whole dataset and figure~\ref{fig:mirWE_syst} shows the statistical errors and the statistical plus systematic errors on the modulus and phase of the calibration transfers from the WE photon calibrator to the WE mirror actuator.
The statistical and systematic uncertainties on the phase are very similar to the one estimated for NE data.
The systematic uncertainties on the modulus are much larger, as expected due to the WE PCal power calibration variations: above 20~Hz, systematic errors are at the level of $0.6\%$ on the whole frequency band.\\
\lr{
The conclusion of this analysis is that the WE mirror actuator response was stable in time with systematic uncertainties smaller than $0.6\%$ in amplitude and smaller than $4~$mrad in phase in the 20~Hz to 1~kHz range,
but that part of these uncertainties comes from the PCal calibration variations, and is not real variations of the WE mirror actuator response.
We do not expect larger time variations at higher frequencies where the weekly data are anyway compatible within their larger statistical uncertainties.
}

Note that independent measurements of the WE mirror actuator based on the free swinging Michelson technique (described later) do not show up any variation in the WE mirror actuator larger than 0.3\%, which confirms that the large variations comes mainly from the PCal.
However, we have chosen to use conservative uncertainties of 0.6\% in modulus 
for the WE mirror actuator calibration (see section~\ref{sec:calibuncertainties}).


\subsubsection{\bf Comparison of WE mirror final O3 model with the online models}

Despite these larger than expected variations, we have averaged all the dataset and extracted a full model for WE mirror actuator. The model parameters and fit uncertainties are given in the last column of table~\ref{tab:mirWE_models}. The fit residuals are within 0.2\% and 2~mrad between 10~Hz and 1~kHz,
and contrary to the NE case, no unexpected behavior is seen on the phase at low frequency.\\

As for NE, the WE actuator models were extracted in March~2019 and in October~2019 with 
the dataset available at that time and they have been used online during O3a and O3b respectively.
The model parameters are summarized in table~\ref{tab:mirWE_models}.
The need to add a high-frequency zero showed up also, which was not included in the O3a online model. The O3b model shows a general offset of 0.5\% compared to the full model, but is in agreement within better than 0.7\% and $1$~\mus\ in the whole band. Due to the absence of a high-frequency zero in the O3a online model, differences in modulus are within 1\% up to 600~Hz, and reach 2.5\% at 1~kHz.

\begin{table}[!ht]
\centering
\begin{tabular}{|c|c|c|c|}
\hline 
Model & O3a online &  O3b online & preO3+O3a+O3b \\ 
\hline
Fit     &  &  &  \\
Gain ($\upmu$m$/$V) & $0.4007$ & $0.399$  &  $0.4012\pm0.0001$ \\  
Pole $f_{p}$ (Hz)   & $113.04$ & $109$    &  $109.9\pm4.4$ \\ 
Zero $f_{z}$ (Hz)   & $115.22$ & $111$    &  $111.7\pm4.5$ \\ 
Zero $f_{z}$ (Hz)   & -        & $5364$   &  $5376 \pm 135$ \\
\hline
Delay ($\upmu$s)   & $-171.2$  & $-142.0$ &  $-142.8\pm0.7$\\
\hline
Pendulum &  &  &     \\
$f_{0}$ (Hz) & $0.6$  & $0.6$  & $0.6$ \\
$Q_{0}$      & $1000$ & $1000$ & $1000$ \\
\hline
\end{tabular}
\caption{\label{tab:mirWE_models} WE mirror actuators models during~O3. 
The models "O3a online" were derived in March 2019 and used online during O3a ;
the models "O3b online" were derived in October 2019 using all the O3a data and were used online during O3b ;
the models "preO3+O3a+O3b" (full model) were derived after the end of O3. 
Their validity range is from $10~$Hz to $1500~$Hz.
}
\end{table}

\subsubsection{\bf Permanent monitoring of mirror actuators during O3}

Over all the O3 run, permanent lines have been injected on NE and WE mirrors through their electromagnetic actuators. An online process (TFMoni) has been setup to compute transfer functions and extract their modulus and phase at the frequency of the permanent lines to monitor their stability~\cite{bib:TFMoni_doc,bib:2021_TFMoni_O3}.
\dv{Each transfer function was computed every 5 seconds as a moving average of 12~FFTs (Fast Fourier Transforms), each computed over 10~seconds of data}.
\msa{Two different transfer functions were computed to monitor the stability of the mirror actuators:}
\begin{itemize}
    \item from the excitation channel generated on a real-time \lr{ computer ($zCAL_i$ in figure~\ref{fig:itfconfig})} to the excitation channel received and sent back to data acquisition by the digital part of the mirror actuator \lr{($zC_i$)}. The aim is to monitor the data transmission and processing in the digital part.
    \item from the excitation channel to the four channels sensing the current flowing in the four coils of the actuator. The aim is to monitor also the analog part of the actuators.
    Variations of less than 0.5\% on modulus and less than 4~mrad on phase were found around 60~Hz, dominated by statistical fluctuations
    (and significant time variations but at the level of~0.04\% and 0.1~mrad around 355~Hz).
\end{itemize}
The computed values were available online and used by a low-latency process~\cite{bib:DMS}~\dv{(called Detector Monitoring System)} to display flags in the control room and trigger alerts in case of unexpected values. 
The stability of the values were monitored at the 0.5\% level in modulus and 4~mrad in phase, dominated by statistical fluctuations.

\subsection{Calibration of the NE and WE marionette actuators}

The end marionette actuators $A_{mar,i}^{Sc}$ , $i = \{$WE,NE$\}$, are also calibrated with the interferometer in its standard operating mode. The technique is similar as for the calibration of the end mirror actuators except that we compare the end mirror motion induced by photon calibrator with the end mirror motion driven by the electromagnetic actuators of the marionette.

\begin{figure}[!ht]
    \begin{minipage}[t]{0.49\textwidth}
	   \hspace{-1.5cm}    \includegraphics[width=0.99\linewidth]{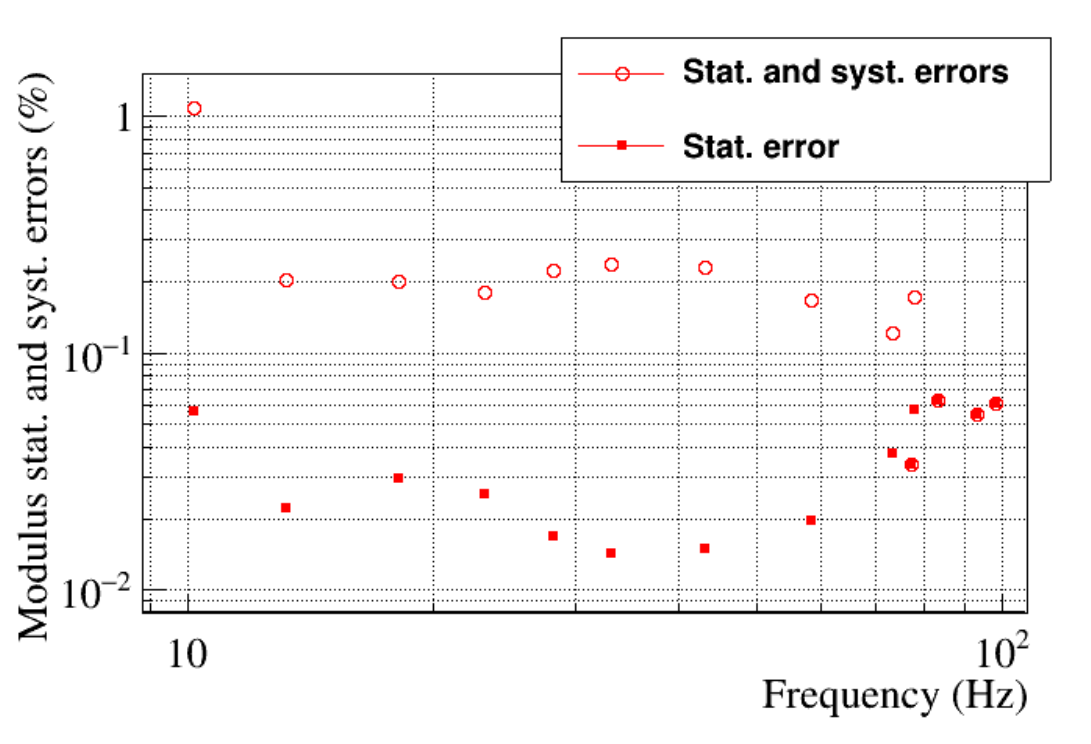} 
    \end{minipage}
    \begin{minipage}[t]{0.49\textwidth}
        \centering
	    \includegraphics[width=0.99\linewidth]{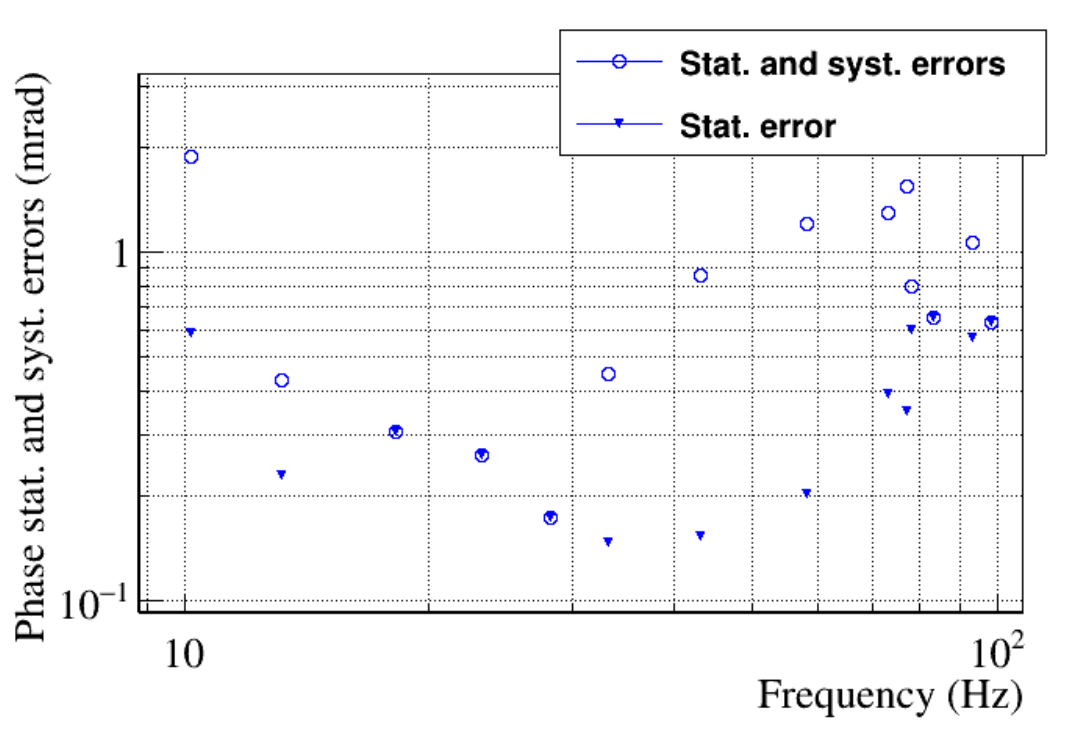}
    \end{minipage}
    \caption{Statistical and systematic errors estimation on the calibration transfer from NE PCal to NE marionette actuator. The errors on the amplitude are expressed in [$\%$] and the errors on the phase are expressed in [mrad]. Left: Statistical errors on the amplitude (red filled squares) of the calibration transfer as a function of frequency and statistical plus systematic errors (red empty circles) estimation to have a p-value$~>0.05$. Right: Statistical errors on the phase (blue filled triangles) of the calibration transfer as a function of frequency and statistical plus systematic errors (blue empty circles) estimation to have a p-value$~>0.05$.}
    \label{fig:marNE_syst}
\end{figure}

As for the end mirrors actuators and shown figure~\ref{fig:marNE_syst}, the stability of these measurements has been checked during O3.
Systematic uncertainties are estimated of the order of 0.2\% and 2~mrad in the 10~Hz-100~Hz band.\\

\begin{figure}[!ht]
    \centering
	\includegraphics[trim={0 0cm 0 0cm},clip,width=0.8\linewidth]{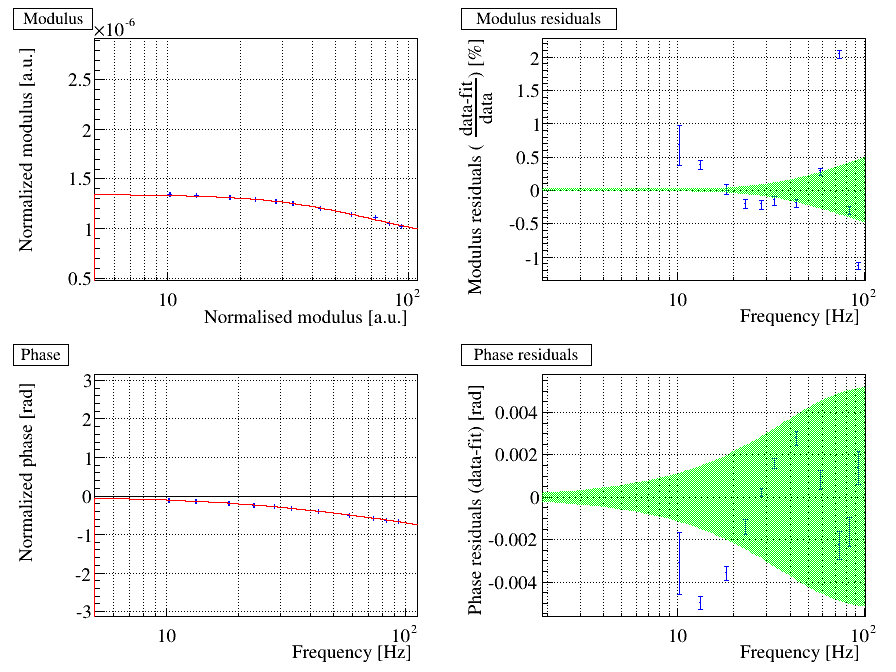} 
    \caption{NE marionette actuator response normalized by the double pendulum mechanical model. On the left, the measurements are represented in blue and the fit is drawn in red. Modulus is given in~m/V and phase in~rad. The modulus and the phase are fitted simultaneously. 
    On the right, the residuals are shown.
    The green area represents the uncertainties given back by the fitted model,
    taking into account parameter covariance.
    }
    \label{fig:marNE_fullmodel}
\end{figure}

The final actuator model results from the average of all the weekly measurements done during~O3, and includes the pre-O3 measurements from March~2019. Figure~\ref{fig:marNE_fullmodel} shows the final NE marionette actuator response and its fit, normalized by a mechanical response model. The mechanical model used for the marionette is made of two poles at 0.6~Hz with a quality factor of~1000, to hide the $f^{-4}$ mechanical attenuation in the plots.
Above 20~Hz, the fit residuals are within 2\% in modulus and 3~mrad in phase.\\

\subsection{Calibration of WI, NI, BS and PR actuators}
\label{sec:calibniwibspr}

Once the actuators response of NE and WE mirrors have been calibrated with the photon calibrators it is possible to calibrate the actuators response of WI, NI, BS, PR mirrors and WI, NI and BS marionettes by transferring the calibration using different interferometer configurations.


\subsubsection{\bf Calibration of NI and WI mirror actuators}
\label{subsec:calibniwi}
The calibration transfer from end to input mirror actuator is done
with the interferometer in its standard working point, but with
the NI and WI mirror actuators switched on (i.e. the DAC output
connected the coil drivers). The input mirror actuators are
not used in observing mode because they are noisy:
they largely spoil the sensitivity below $\sim200$~Hz. 
\lr{In this configuration, sinewave excitations are sent to the end mirrors ($Exc_{ref}$) and then to the input mirrors ($Exc_{new}$),  and the interferometer main photodiode output $\mathcal{P}_{DC}$ is used as output channel ($S_{out}$ in equation~\ref{eqn:transfer_tfratio})}. 
Compared to the description given in section~\ref{sec:calibStrategy}, the optical response $R$ is not exactly the same in both dataset and does not cancel when doing the transfer function ratio (see equations~\ref{eq:transfer_Sref}, \ref{eq:transfer_Snew} and~\ref{eqn:transfer_tfratio}): hence some corrections must be applied to properly take the difference into account. 
A motion of an end mirror couples only to the differential arm length degree of freedom of the interferometer, while a motion of an input mirror also couples to the short Michelson degree of freedom (see section~\ref{sec:detector}). Since the finesse of the arm Fabry-Perot cavities is high, of the order of~450, the coupling to differential arm length dominates and the difference is small: the interferometer response to a motion of an end mirror has an amplitude greater by 0.37\% than the response to a motion of an input mirror. On the phase, the response to an end mirror motion has a higher delay than the response to the motion of an input mirror, by 10~\mus, due to the light propagation time in the arm cavity. \\

\lr{Measurements in this special configuration} were done every week during O3. The ratio of the transfer functions was computed on every dataset and averaged over the whole run.
The statistical and systematic uncertainties on this ratio used for the transfer from NE to NI (WE to WI) mirror actuator calibration 
are below  0.3\% and 1~mrad in modulus and phase above 20~Hz. \\

The measured response of the input mirror actuator is then computed multiplying the data points of the averaged end mirror actuator by the data points of the averaged transfer function ratio. 
It is then fitted and the fitted gain and delay are finally corrected for the 0.37\% and 10~$\mu$s described earlier. The response is modeled by a simple pole around 320~Hz, consistent with the response expected for this actuator in strong dynamic/high noise configuration.

\subsubsection{\bf Calibration of BS and PR mirror actuators }

\lr{The calibrated WI mirror actuator is then used as reference to calibrate the BS and PR mirror actuators.}
For this measurement, the detector is partially misaligned and only the Fabry-Perot cavity made of
PR-BS-WI mirrors is aligned and locked. 
In this configuration, sine wave excitations are sent to the WI mirror ($Exc_{ref}$) and then to the PR and BS mirrors ($Exc_{new}$), while the output channel $S_{out}$ is extracted from the auxiliary photodiode used to control the cavity lock, sensitive to the power inside this cavity. \\

The systematic uncertainties are estimated from the weekly measurements, in the range 20~Hz and 500~Hz: 
they are lower than 0.6\% in modulus and 6~mrad in phase.\\

The measured responses of the BS and PR mirror actuators are then computed multiplying the data points of the WI mirror actuator by the data points of the averaged transfer function ratio. 
These new data are fit 
and the fitted gain and delay are finally corrected for the 0.37\% and 10~\mus\ described earlier. 

\subsubsection{\bf Calibration of marionette actuators (BS, NI, WI) }
The BS marionette was controlled continuously during O3.
The NI and WI marionette were controlled only in the so-called "Earth Quake mode" of the interferometer, when the longitudinal control of the arms differential length was done 
using the NI and WI marionette actuators in order to be less sensitive to large seismic motions 
induced by earthquakes or bad weather.

As a consequence, the controls were subtracted in the $h(t)$ strain reconstruction processing.
The actuators of the BS, NI and WI marionette were hence calibrated,  via different transfers:
from BS mirror to BS marionette, and from NE, WE PCal to NI and WI marionette.

\subsection{Detector calibration uncertainties estimation}
\label{sec:calibuncertainties}

Each step of the calibration procedure described in the previous sections contributes 
to the total uncertainty of the actuator response both in amplitude and in phase. 
Tables~\ref{tab:MirUncertainties} and~\ref{tab:MarUncertainties}
provide the total uncertainty for the mirror and marionette actuators respectively.
In the case of the mirror actuators, the breakdown of the various contributions to the total uncertainty described in previous sections is included. 

\begin{table}[tb]
\begin{center}
\small{
\begin{tabular}{|c|c|c|c|c|c|}
\cline{3-6} 
\multicolumn{2}{ c|}{}                          & NE mirror       & WE mirror        &  BS mirror &  PR mirror \\
\hline
\multicolumn{2}{|c|}{Stat. uncertainty}         & 0.2\% (6 mrad)  & 0.2\% (2 mrad)   &  1\% (10 mrad) & 2\% (50 mrad)  \\
\multicolumn{2}{|c|}{and fit residuals}         &                 &                  &                &                \\
\hline 
\multirow{5}[0]{*}{\rot{Syst. uncert.}}   
 & PCal calibration                             & 1.39\% (0 mrad) & \multicolumn{3}{c|}{1.73\% (0 mrad)} \\
 & PCal to end transfer                         & 0.3\% (3 mrad)  & \multicolumn{3}{c|}{0.6\% (4 mrad)}  \\ 
 \cline{2-6}
 & WE to WI transfer                            & --              & --               & \multicolumn{2}{c|}{0.2\% (1 mrad)}  \\ 
 & WI to BS transfer                            & --              & --               & 0.6\% (6 mrad) &  --                  \\
 & WI to PR transfer                            & --              & --               & --             & 0.6\% (6 mrad)       \\
 \cline{2-6}
 & PCal readout delay                           & \multicolumn{4}{|c|}{3~\mus}  \\
\hline
\multicolumn{2}{|c|}{Total uncertainty}         & 1.44\%           & 1.84\%           & 2.18\%        & 2.78\%      \\
\multicolumn{2}{|c|}{(quadratic sum)}           & 6.7 mrad         & 4.5 mrad         & 13 mrad       & 12.4 mrad     \\
\multicolumn{2}{|c|}{}                          & 3~\mus           & 3~\mus           & 3~\mus        & 3~\mus      \\
\hline                                                                                                            
\multicolumn{2}{|c|}{Validity range}            & 20-1500 Hz        & 20-1500 Hz      & 20-500 Hz     & 20-500 Hz   \\
\hline
\end{tabular}
\caption{Summary of the statistical and systematic uncertainties on the mirror actuator models.
  The uncertainty on the modulus and phase is given for each contribution to the total uncertainty. The sum of all the contributions to each mirror actuator model, and its validity range, is provided. \lr{ The validity range does not span the full range of strain frequencies but are enough given the unity gain frequency of the different loops acting on the mirror positions. }
  See text for details.
}
\label{tab:MirUncertainties}
}
\end{center}
\end{table}

\begin{table}[tb]
\begin{center}
\small{
\begin{tabular}{|c|c|c|c|c|c|c|}
\cline{3-7}
\multicolumn{2}{c|}{}                          & NE mario.         & WE mario.        & BS mario.      & NI mario. & WI mario. \\   
\hline                                                                                                 
\multicolumn{2}{|c|}{Total uncertainty}        & 2.0\%         &   1.9\%             & 2.7\%           & 3.3\%      & 3.5\%    \\ 
\multicolumn{2}{|c|}{(quadratric sum)}         & 11 mrad       &   8 mrad            & 15 mrad         & 30 mrad    & 30 mrad  \\       
\multicolumn{2}{|c|}{}                         & 3~\mus        &   3~\mus            & 3~\mus          & 3~\mus     & 3~\mus    \\       
\hline                                      
\multicolumn{2}{|c|}{Validity range}           & 10-100 Hz     &   10-100 Hz         & 10-60 Hz       & 10-80 Hz   & 10-80 Hz  \\                  
\hline                                                                               
\end{tabular}
\caption{Summary of the uncertainties on the marionette actuator models, along with their validity range.
    \lr{ The validity range does not span the full range of strain frequencies but are enough given the unity gain frequency of the different loops acting on the mirror positions. }
  See text for details.
}
\label{tab:MarUncertainties}
}
\end{center}
\end{table}

\section{Comparison with calibration using free swinging Michelson technique}
\label{sec:calib_michelson_compare}
Free swinging Michelson technique is an independent method to calibrate the mirror
electromagnetic actuators, using the laser wavelength (1064~nm) as length \lr{standard}.
It has been the reference method for calibrating Virgo until \lr{the end of the O2~run}. 
During~O3, it has been used as an independent cross-check of the reference
calibration based on the PCal technique described in the previous section.

Free swinging Michelson calibration during~O3 is similar to the~O2 calibration described in~\cite{bib:2018CQGra..35t5004A,2011CQGra..28b5005A}.
The interferometer mirrors are either aligned or misaligned to setup a free swinging short Michelson. 
The differential arm length $\Delta L(t)$ is measured using a non-linear reconstruction (described in~\cite{2011CQGra..28b5005A})
from the interference fringes passing on the output photodiode. The NI, WI and BS mirror actuator response, in meter per volt, is estimated by applying known excitations to the mirror electromagnetic actuators and observing their effect on the reconstructed $\Delta L$.
Then, calibration transfers from the input NI and WI mirror actuator to the end NE and WE mirror actuators are done
(similarly as described in section~\ref{subsec:calibniwi}): with the full interferometer locked, one can compare the effect of known motions of the NI and WI mirrors on the dark fringe power to the effect of known excitation of the NE and WE mirrors. This comparison allows to estimate the NE and WE mirror actuator responses, based on the free swinging Michelson technique.\\

After an overview of the Advanced Virgo photodiode readout and the limitations of this technique,
we compare its results with the PCal-based calibration results.

\subsection{Note on the Advanced Virgo photodiode readout }

\lr{
The output beam used to observe the interference fringes is a small fraction of the interferometer output beam
extracted before the output mode-cleaner cavity~\cite{bib:paperOMC}. It is sent onto a photodiode,
and the non-linear $\Delta L$ reconstruction uses two channels from this photodiode:
}
$\mathcal{P}_{DC}$, that measures the output power components from 0 to 10~kHz,
and $\mathcal{P}_{AC}$, that is extracted using digital demodulation at 56~MHz.
The $\mathcal{P}_{DC}$ signal is in practice the blending of two output channels: 
$\mathcal{P}_{DC,raw}$ channel from DC to 10~kHz with high dynamic and 
$\mathcal{P}_{Audio}$ channel, \lr{high-passed at $f_0$ ($\sim5\,\text{Hz}$), with less noise but lower dynamic, in the band 5~Hz to 10~kHz.}
We discovered a coupling between both channels: the response of the $\mathcal{P}_{DC,raw}$ channel is modified because of the presence of the high-pass analog filter in the audio channel.

Two different photodiodes receive respectively 90\% (PD1) and 10\% (PD2) of the extracted beam power.
Since there is less power on it, the PD2 photodiode provides less sensitive calibration than PD1.
The Virgo laser power was increased by about a factor two between O2 and O3~runs. 
To prevent saturation of the $\mathcal{P}_{Audio}$ channel, the PD1 photodiode readout electronics was modified:
the frequency of the audio-filter was increased from 5~Hz to 16~Hz. 

\lr{
In order to measure the responses of the $\mathcal{P}_{DC,raw}$ channels,
a LED was put manually in front of the photodiode,
and driven with a noise to modulate its power~\cite{bib:2019_vnote_calibTiming}.
This measurement could be done only during the rare periods 
when the vacuum tank hosting the optical bench was opened.
Hence, it was not monitored during the run.
}
The response was fit with a pole and a zero around the high-pass filter frequency $f_0$. 
However, doing the measurement with the photodiode temperature going from $54^\circ\text{C}$ (few minutes after opening the vacuum tank, during thermal transient) down to $38^\circ\text{C}$ (steady temperature when the optical bench is in air) showed that the pole and zero frequencies vary with temperature.
The correlation could not be measured precisely, but whitening filters have been derived\footnote{
fixed pole and zero frequencies, extrapolated at temperatures close to the photodiode temperature measured during free swinging Michelson measurements.
}
and then applied in real-time on the $\mathcal{P}_{DC,raw}$ channel, to provide a channel $\mathcal{P}_{DC,whitened}$ with flat response from DC to 10~kHz. Then, the final channel $\mathcal{P}_{DC}$ during~O3 has been computed by blending $\mathcal{P}_{DC,whitened}$ and $\mathcal{P}_{Audio}$, so that in principle the absolute calibration is correct and no frequency-dependent error is introduced by the photodiode readout.

\subsection{Limitations of free swinging Michelson during O3}

One of the main limitation of the free swinging Michelson technique comes from the issue of non-flat DC response
described above: (i) the whitening filter does not perfectly compensate for the readout non-flatness,
and (ii) the compensation must slightly vary in time for PD1 since its temperature increases by about $0.3^{\circ}\text{C}$ during the free swinging Michelson measurements, after a shutter has been opened
and the laser beam reaches the photodiode.
Using the most sensitive photodiode, PD1, the measurement of the mirror actuation is indeed distorted by 1\% in the range  10~Hz to 100~Hz because the whitening filter does not match perfectly the real response. But using the less sensitive photodiode, PD2, the measurements have statistical uncertainties of the order of 1\%.
In addition, comparing both results, the mirror actuation response when using PD1 is 1\% to 2\% higher than when using PD2, showing the level of systematic errors.

Another limitation of this technique is the limited $\Delta L$ sensitivity of the Michelson configuration,
of the order of $10^{-13}\,\text{m/}\sqrt{\text{Hz}}$. Large NI and WI mirror motions must be applied to
calibrate their actuators, few order of magnitudes larger than the mirror motions that are controlled in
the standard condition of the interferometer. In particular, the input mirror motions applied for the
calibration transfer to the end mirrors are five orders of magnitude lower. Hence, linearity of the actuation response is a strong assumption of this technique. 

Finally, this technique, based on the Virgo laser wavelength, cannot be intercalibrated with other detectors.
The possibility to cross-calibrate the LIGO and Virgo photon calibrators is one very important advantage of the
PCal technique.

\subsection{Comparison with the reference calibration}
\begin{figure}[!ht]
    \centering
	\includegraphics[trim={0 0cm 0 0cm},clip,width=0.8\linewidth]{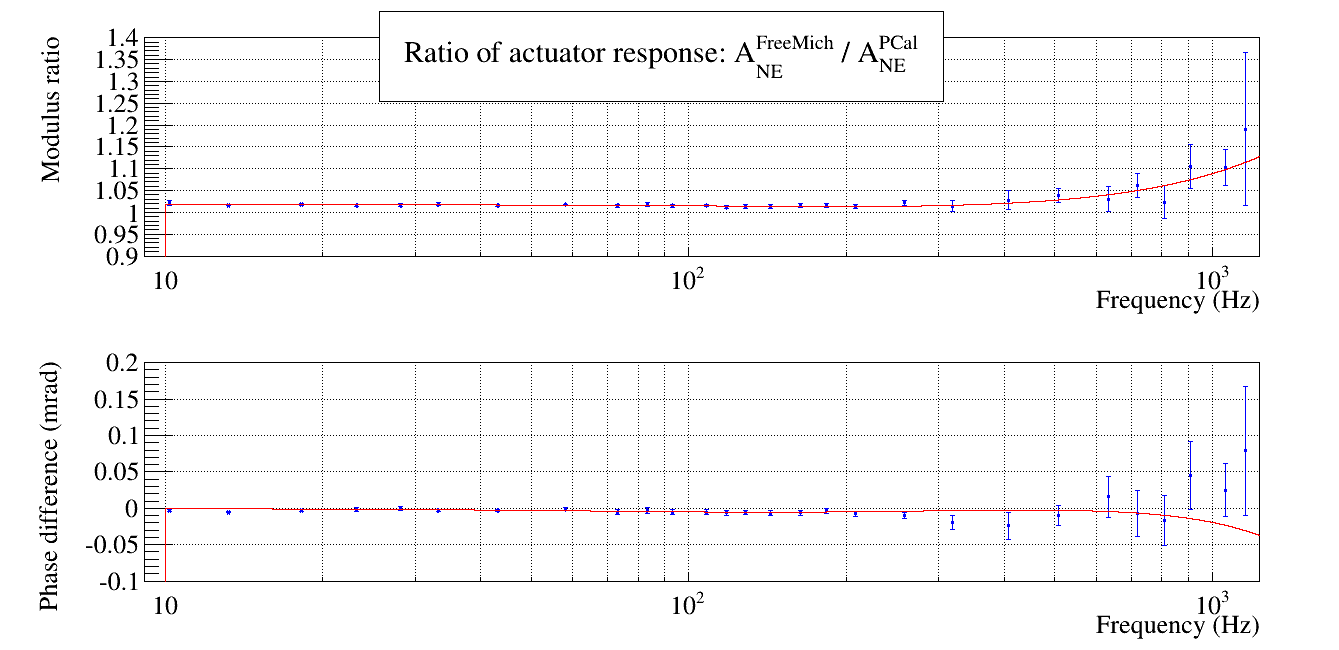} 
    \caption{
    Ratio of the NE mirror actuator response measured with the free swinging Michelson technique 
    to the response measured with the PCal technique.
    The blue points give the ratio of the actuator response data points measured with both techniques,
    with their associated statistical uncertainties.
    The red curve shows the ratio of the models extracted from the two independent techniques.}
    \label{fig:mirNE_FreeMichVsPcal}
\end{figure}

Figure~\ref{fig:mirNE_FreeMichVsPcal} shows the ratio of 
the NE mirror actuator response fit on measurements based on free swinging Michelson technique 
to the response fit on measurements with the PCal technique.
The ratio of the data points measured using both techniques are shown,
along with the ratio of the two fitted models.
The statistical errors are dominated by the free swinging Michelson measurements at all frequencies. 
A systematic difference of about 1.5\% is seen in modulus
(the mirror actuation response, in m/V, being larger using the free swinging Michelson technique),
indicating the level of agreement of the absolute calibration using both techniques.
This difference is flat within better than 0.7\% between 10~Hz and 400~Hz, where the actuation response modulus
varies by 4\% (see figure~\ref{fig:mirNE_fullmodel}): it confirms the presence of the pole and zero around 120~Hz in the actuator response.
At high frequency, above about 500~Hz, there is a slight trend of increasing difference between the two kinds of measurements, but it is not significant within the large statistical uncertainties of free swinging Michelson data. 
The phase measurements agree between both techniques. At high frequency, the two models diverge following a difference of  about 4~\mus\ in delay, but again the data themselves show that this difference is not significant.\\

These small differences are compatible with the uncertainties estimated from both methods on the NE mirror actuation:
1.44\% using the PCal as reference and between 1\% and 2\% using the free swinging Michelson technique.
We conclude that the cross-check with the free swinging Michelson technique validates the PCal-based calibration within better than 2\%.

%% file: hrec.tex
\newpage
\section{Reconstruction of the detector strain h(t) and noise subtraction}
\label{sec:hrecAndNoiseSubtraction}
The Virgo detector strain time series $h(t)$ is reconstructed using the raw detector time series
and the models describing the sensing chain and the controlled mirror actuators whose calibration has been
described in the previous sections. 
In this section, we report the principle of this $h(t)$ reconstruction
\lr{and the noise subtraction used during O3 which was applied within the same process. 
After a general introduction on the overall processing,
more details are given on the $(t)$ reconstruction itself in section~\ref{sec:hrec},
and then further details on the noise subtraction and noise-witness channels considered during O3 are provided} in section~\ref{sec:NoiseSubtraction}. 
In section~\ref{sec:hrecerrors}, we then describe the method used
to estimate the uncertainties on $h(t)$ 
and we give them for both O3a and O3b periods.

\subsection{Principle}
\begin{figure}[!ht]
    \centering
	\includegraphics[trim={0 0cm 0 0cm},clip,scale=0.35]{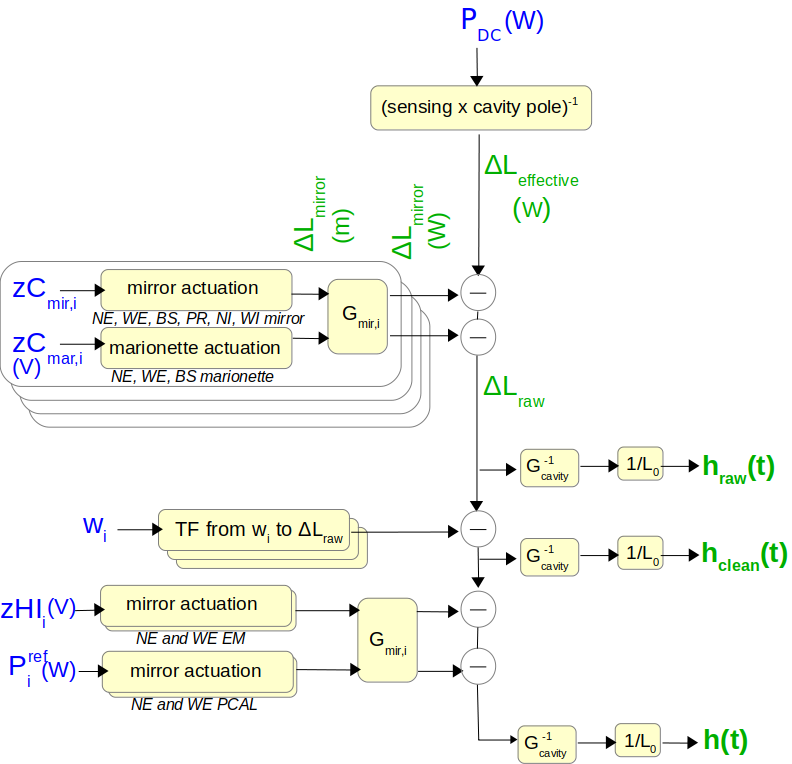} 
    \caption{
    Principle of the Virgo $h(t)$ strain data reconstruction and noise subtraction during O3~run,
    with different steps: 
    the reconstruction itself, producing the $h_{raw}$ intermediate channel,
    the noise subtraction, producing the $h_{clean}$ intermediate channel
    and the subtraction of the hardware injections, producing the final $h(t)$ channel.
    Input data channels are shown in blue, output data channels in green. 
    \lr{
    P$_{DC}$ is the main output power of the interferometer.
    $zC_i$ is the control signal applied to the mirror or marionette electromagnetic actuator.
    $G_i$ is the gain of the optical response of the interferometer to a motion of mirror~$i$.
    $L_0 = 3\,\text{km}$ is length of the arm cavities.
    $w_i$ are the noise witness channels used for noise subtraction.
    The NE and WE mirror motions applied via the hardware injections are estimated and subtracted
    using the channels used as reference for the actuator calibration:
    $zHI_i$ for the mirror electromagnetic actuators and
    P$^{ref}_i$ for the PCal actuators. 
    }
    }
    \label{fig:hrecprinciple}
\end{figure}

In the first steps of the $h(t)$ reconstruction, the contributions of the control signals are removed from the dark fringe signal,
taking into account the interferometer optical transfer function, as sketched in figure~\ref{fig:hrecprinciple}.
Following the notations of figure~\ref{fig:itfconfig}, inputs are the measured dark fringe photodiode channel, \msa{$\mathcal{P}_{DC}$}, the \msa{control signals}, $zC_i$, sent to the mirror and marionette actuators,
and the calibrated transfer functions for the photodiode readout and the different actuators.\\

\begin{figure}[!ht]
    \centering
	\includegraphics[trim={0 0cm 0 0cm},clip,width=0.6\linewidth]{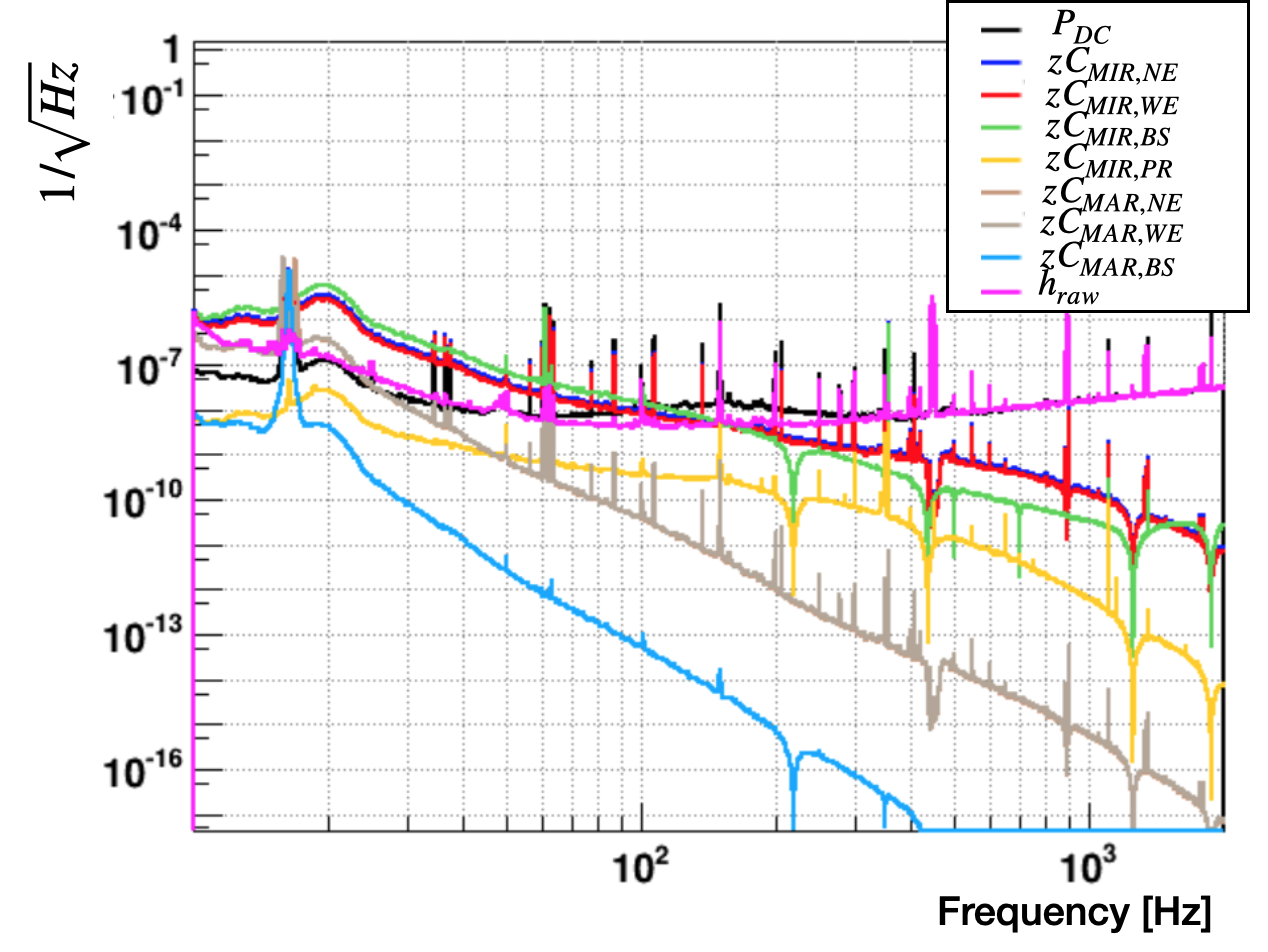} 
    \caption{
    Contributions of the mirrors longitudinal controls
    to the reconstructed signal~$h_{raw}$ in O. Black curve is the
    amplitude spectral density of the dark fringe signal~$\mathcal{P}_{DC}$. \msa{Pink} curve is the amplitude spectral density of $h_{raw}$ purple in watt units. The other curves are the contribution of the various controls $zC_i$ in the same units.}
    \label{fig:hrec_breakdown}
\end{figure}

The controls subtracted from the dark fringe signal to obtain~$h_{raw}(t)$ have contribution mainly at low frequency and
have almost no effect on the amplitude of~$h_{raw}(t)$ above 300~Hz
as can be seen in figure~\ref{fig:hrec_breakdown}.\\

The computation is done in the frequency domain using fast Fourier transforms over 8~s with 4~s overlap.
Calibration lines, added to the control signals~$zC_i$, are used to monitor the time varying parameters of the optical response, i.e. the mean optical gain and mean cavity finesse. 
These parameters slowly vary, with the interferometer alignment for instance.
The monitored values are modified continuously in the $h_{raw}(t)$ reconstruction processing. \\

Once the raw time series~$h_{raw}(t)$ is reconstructed, noise subtraction is applied to get
the cleaned signal~$h_{clean}(t)$.
During the commissioning of the Virgo interferometer, various noise contributions were still
present in the dark fringe signal and could be subtracted during O3 thanks to their presence also in auxiliary monitoring channels, which could be used as noise witnesses. 
Finally, the hardware injections (both continuous lines and pulsar-like signals) are also subtracted,
using the excitation channels and the actuator calibrated models, to get the final strain time series $h(t)$.

The $h(t)$ reconstruction and noise subtraction processes being done in the frequency domain, 
the final $h(t)$ channels are obtained at 20000~Hz and 16384~Hz by performing inverse Fast Fourier Transforms. 
The $h(t)$ time series computed online and provided publicly
in GWOSC~\cite{bib:GWOSC} is called V1:Hrec\_hoft\_16384Hz.

\newpage

\subsection{Main changes with respect to O2}

\msa{The $h(t)$ reconstruction algorithm, which includes noise subtraction, was} not changed between the code version
used to reprocess O2~data and the code version used online during the O3~run. The main update in the reconstruction itself concerns the reduction of the latency of the data processing.
Using FFTs of 8~s, instead of 20~s as during O2, reduced the latency introduced by the reconstruction
from 20~s to 8~s, which is an important improvement to provide low-latency public alerts
in case of gravitational wave detections. 

The noise subtraction algorithm was initially developed during~O2 to subtract a single witness-channel that monitored the frequency noise.
During~O3, five different witness channels were identified and used to subtract different sources of noise.
These are described later, as well as the limitations of the method when different witness channels are correlated.

Twelve continuous hardware injections of sinusoidal signals were performed during~O3 in the
detector most sensitive frequency band. The goal of these injections \msa{is to monitor the time variation
of a systematic bias in the amplitude and phase of the reconstructed~$h(t)$ signal.}
Such a frequency-dependent bias was seen during~O2 (and in earlier Virgo science runs),
but was monitored only weekly. Such injections have allowed to assess the stability of this bias
during~O3 (as shown in section~\ref{sec:hrecerrors}).


\subsection{h(t) reconstruction}
\label{sec:hrec}

\subsubsection{Monitoring of time varying parameters: optical gains and cavity finesse}
\begin{figure}[!ht]
    \centering
	\includegraphics[trim={0 0cm 0 0cm},clip,scale=0.4]{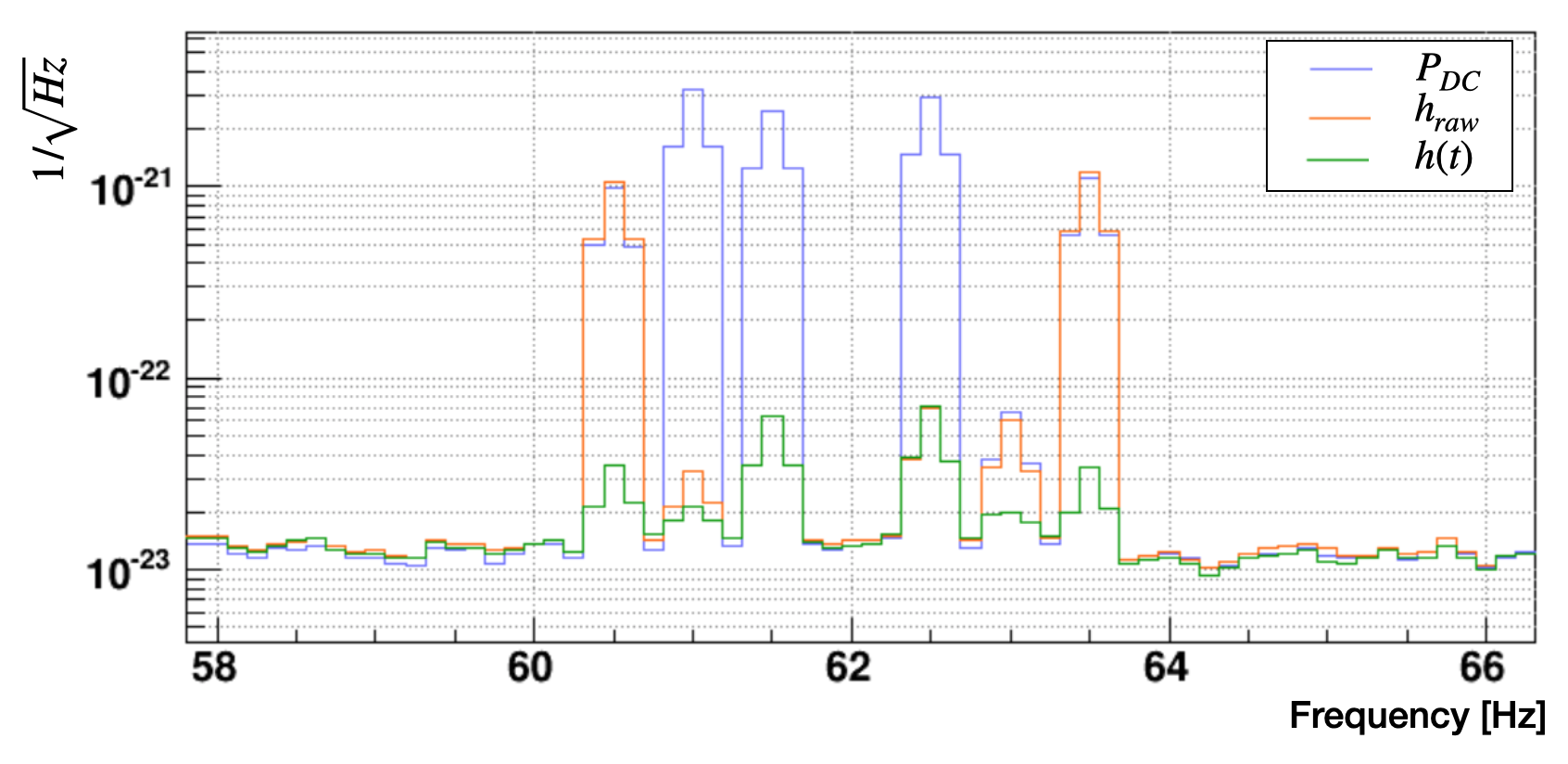} 
    \caption{Calibration lines in the spectrum of $\mathcal{P}_{DC}$, before $h(t)$ reconstruction (blue), $h_{raw}$ (orange) and the reconstructed $h(t)$(green), delivered for data analysis. 
    \dv{P$_{DC}$ has been normalized so that it is at the same mean level as $h_{raw}$ between 56~Hz and 58~Hz.
    The four calibration lines used for the $h(t)$ reconstruction are visible in $\mathcal{P}_{DC}$ and
    well reduced in $h_{raw}$ (and $h(t)$) as expected.
    The two lines at 60.5~Hz and 63.5~Hz are hardware injections from the PCal: 
    they are also well reduced in the final $h(t)$ channel.}
    }
    \label{fig:callinesO3}
\end{figure}

In order to monitor the time varying parameters of the detector optical response,
permanent lines in the form of sinusoidal signals were injected on WE, NE, PR and BS mirror electromagnetic actuators
via the channels $zCAL_i$ (see figure~\ref{fig:itfconfig}) around 60~Hz as summarized in table~\ref{tab:callines}.
The lines were injected with a signal-to-noise ratio of the order of~100 above the detector background noise
as summarized in the table~\ref{tab:callines}, and naturally subtracted in the $h(t)$ reconstruction
algorithm since they were included in the control signals $zC_i$. 
Note the coupling of the motion of the PR mirror to the detector output is low and varies
with the detector working conditions.
As a consequence, the signal-to-noise ratio for the PR mirror induced motion is of the order of a few, 
with significant variations.
From these lines, the optical gains $G_i$ and the cavity finesse (used in the model of the cavity pole shown in figure~\ref{fig:hrecprinciple}) were updated at the pace of the $h(t)$ reconstruction, that is once every 4~s. 
Figure~\ref{fig:callinesO3} shows the calibration lines in the spectrum of the dark fringe signal, and in the spectrum of the reconstructed $h(t)$ before and after their mitigation.

\begin{table}[!ht]
\centering
\begin{tabular}{|c|c|c|c|}
\hline 
Injection source & Line frequency &  Line SNR \\ 
\hline
NE mirror actuator &  62.5 Hz  &  120  \\
WE mirror actuator &  61.5 Hz  &  100  \\
BS mirror actuator &  61.0 Hz  &  120  \\
PR mirror actuator &  63.0 Hz  &  3  \\
\hline
\end{tabular}
\caption{\label{tab:callines} Permanent calibration lines injected during~O3 to monitor the optical gain and finesse of the arm cavities.
}
\end{table}

\subsubsection{Notes on violin modes not modeled in actuator responses}

The various spectral lines around 450~Hz, due to the mirror suspensions violin modes have been well identified and associated to each mirror.
In \msa{ observing mode, which corresponds to the periods where the detector is stable enough and data are exploitable for astrophysical purposes,} the mirror control signals were notched at these frequencies to prevent exciting the violin modes. 
These violin modes were not included in the mirror actuation models used during~O3.
As a consequence, the reconstructed $h(t)$ strain channel must be biased at these frequencies. However, the detector sensitivity is spoiled at these frequencies (the violin modes, between 450~Hz and 460~Hz, are more than three orders of magnitude above the Advanced Virgo sensitivity) and the data are thus not useful for any astrophysical analysis.


\subsubsection{Notes on optical response at low frequency}

The optical response of the Virgo power-recycled interferometer with Fabry-Perot cavities in the arms 
is expected to follow a single pole transfer function, 
whose pole frequency depends on the finesse of the arm cavities.
This model has been used in the $h(t)$ reconstruction processing, with a nominal pole at 55~Hz,
and the pole frequency variations were monitored using calibration lines as described earlier.

However, it was discovered during~O3 that the real optical response of the Virgo interferometer
showed a different behavior at low frequency: below about~20~Hz, the measured response
showed an unexpected high-pass behavior. It could be fit by a model used to describe
an optical spring effect, expected to appear in a dual-recycled interferometer configuration
and seen in the LIGO detector response~\msa{\cite{viets2018reconstructing}}. 
A possible explanation for the presence of this optical spring in Virgo during O3 is the non-null reflection of the anti-reflective coating of the lens installed at the position of the future signal-recycling mirror of Advanced Virgo. 
In addition, the shape of this optical response at low frequency varies in time. 
No clear correlation with the input laser power (increased between O3a and O3b), 
nor the differential arm length offset (that was sometimes changed during O3) could be found.
Note that there were no data taken in precise conditions, controlling the mentioned parameters,
hence a detailed study could not be done to understand this effect.

As a consequence, the optical response model used in the $h(t)$ reconstruction was 
not precise at low frequency, and time variations were not monitored. 
The expected effect is a time varying bias on the reconstructed strain channel $h(t)$.
It is included in the estimation of the bias and uncertainties described in section~\ref{sec:hrecerrors}.


\subsection{Noise subtraction}\label{sec:NoiseSubtraction}

\subsubsection{Principle}

\msa{
Remaining noises can couple to the differential arm length variation and add undesired contribution to the detector strain. These can degrade the sensitivity of the detector or even mimic gravitational wave signals. Noise-witness channels are used to unveil remaining noises in the reconstructed detector strain $h_{raw}(t)$.}
\msa{Various noise subtraction methods have been investigated and implemented in the case of LIGO~\cite{driggers2019improving, davis2019improving, vajente2020machine}, 
whose results motivated its implementation as a step following the $h_{raw}(t)$ \dv{online} reconstruction in Virgo.
Note that, in Virgo, the noise subtraction computation is done in the same process as the $h(t)$ reconstruction.}

For each \dv{witness channel}, a transfer function \dv{with $h_{raw}$} (TFs) is used in a specific frequency band to quantify the noise contribution that should be subtracted. These TFs are computed every 500~s and are used to do the noise subtraction over the next 500~s. The noise subtraction is performed in the frequency domain, where each subtracted term is the TF multiplied by the FFT of the noise-witness channel. \\

In the noise subtraction implementation used during O3, for a given noise-witness channel, the TF is computed between h$_{1Noise}$ and the noise witness channel. h$_{1Noise}$ is the \textit{cleaned} $h(t)$ from which all the noise-witness signals have been subtracted except the noise-witness channel of interest, in order to isolate each noise channel contribution.
The coherence between h$_{1Noise}$ and the noise-witness channel is also computed over T=500~s and if, in a frequency bin, this coherence is above a certain threshold (4\% in O3), the TF is updated with the new value. Else, the TF is set to 0 in this frequency bin. More details on the method used to perform noise subtraction can be found in~\cite{bib:TDSHrec}.

During O3, the noise subtraction has been done 
\msa{in an iterative way, on a one-by-one basis, for each noise-witness channel and following the same order of table~\ref{tab:noise_sub}}. 
\msa{An issue of this method is that correlations among noise-witness channels are not taken into account. Checks on the coherence between the various noise-witness channels were done during the O3 commissioning run to study if several noise-witness channels were observing the same noise. 
The frequency band at which each noise-witness channel is considered is selected so that the present noise is only removed once. A caveat of this approach is that any changes on the coherence among noise-witness channels due to changes on the interferometer are not automatically \dv{taken into account and, as a consequence,} regular and manual checks on these quantities should be performed to avoid adding noise to $h(t)$. To tackle this issue, a new method consisting on the linear subtraction of coupled noises, based on \cite{davis2019improving}, is being implemented for the next observing run O4.
}

\subsubsection{Selection of noise-witness channels}
Witness channels identified before the O3 observing run with sufficient reliability have been used to monitor the presence of noise in $h_{raw}(t)$ and to subtract them. These noises are summarized in table~\ref{tab:noise_sub}. \\

{\bf Michelson noise:} The motion of the BS mirror creates a differential signal between the two interferometer arms. Compared to the arm cavity mirrors the impact of this motion is reduced by the optical gain of the arm Fabry-Perot cavity. But the Michelson control noise is much higher than the differential arm length control noise. It yields a non-negligible contribution to the overall interferometer noise.
To subtract this noise contribution, the control signal LSC\_MICH is used as witness channel.\\

{\bf Frequency noise:} The two arm cavities of the interferometer are kept in resonance, and if there is a variation
of this state, a change in the phase of the
reflected light induces a change of the interference observed 
\msa{But this change of the interference profile can also be caused by a laser frequency noise in the cavities. 
In principle, the laser frequency noise induces common variations in the North and West arm cavities and its effect may be canceled thanks to the interference condition. Nevertheless, any optical asymmetry of the arm cavities yields to a non-zero contribution. The laser frequency noise is monitored by the B2 photodiode, used as a noise-witness channel, which sees} the laser beam coming back from the interferometer to the input laser.  \\

{\bf 56 MHz RIN noise:} The Output Mode Cleaner cavity (OMC) filters out the non TEM00\footnote{Transverse ElectroMagnetic mode of the laser beam} modes of the output laser beam created by optical defaults and misalignments,
and the radio frequency sidebands used for interferometer control~\cite{bib:paperOMC}.
In practice, a small fraction of these modes and sidebands is transmitted by the OMC and add phase variations in the dark fringe signal. It was found that the largest contribution to these variations comes from the power fluctuations of the auxiliary modulation at $\pm$56 MHz, known as 56 MHz Relative Intensity Noise.  Such contribution can be subtracted from $h(t)$ by using the reflected power of the OMC as a noise witness channel.\\

{\bf Scattered light noise:} Another noise contribution to be considered is the scattered light from the benches in transmission of the NE and WE mirrors, which can re-couple back into the beam of the interferometer arms. This light reflects on optics that move with seismic noise and can thus add phase noise into the arm beams. Photodiodes on both end benches are sensitive to this scattered light and thus used as noise-witness channel. Further details on scattered light can be found in~\cite{bib:was2021end}.\\

The witness channels used for the Michelson noise and the frequency noise were partially coherent in the band 50 to 150~Hz. The noise subtraction process did not take into account this coherence. It was thus needed to subtract those two noises in separate frequency bands. The limit between those two noise was set between 60 and 150~Hz and was changed a few times during O3, as summarized in table~\ref{tab:noise_sub}. The rest of noise-witness channels were not coherent, hence frequency ranges with some overlap could be used.

\begin{table}[!ht]{
\centering
\begin{tabular}{|l|l|c|c|c|}
\hline 
\small{Noise type} & Channel name & From & Until & Freq. Band (Hz)\\ 
\hline
Michelson noise & \small{V1:SPRB\_B4\_56MHz\_Q} & 2019-04-05 & 2019-08-01 & 8-90 \\
& & 2019-08-01 & 2019-11-01 & 8-150 \\\cline{2-5}
& \small{V1:SSFS\_Err\_Q\_unnorm\_10kHz} & 2019-11-01 & 2020-02-04 & 8-85 \\ \cline{2-5}
& \small{V1:LSC\_MICH} & 2020-02-04 & 2020-02-11 & 8-85 \\
& & 2020-02-11 & 2020-04-02 & 8-60 \\
\hline \hline
Frequency noise & \small{V1:SIB2\_B2\_8MHz\_I} &
2019-04-05 & 2019-08-01 & 90-3500 \\
& & 2019-08-01 & 2019-11-01 & 150-3500 \\
& & \textit{\color{blue}{2019-09-16}} & \textit{\color{blue}{2019-09-30}} & \textit{\color{blue}{50-3500}} \\
& & 2019-11-01 & 2019-11-24 & 85-3500 \\ \cline{2-5}
& \small{V1:SIB2\_RFC\_PD2\_Audio} & 2019-11-01 & 2019-11-24 & 100-3000 \\ \cline{2-5}
& \small{V1:SIB2\_B2\_8MHz\_I} & 2019-11-24 & 2019-11-26 & 8-3500 \\
& & 2019-11-26 & 2020-02-11 & 85-3500 \\
& & 2020-02-11 & 2020-04-02 & 60-3500 \\ 
\hline \hline
56 MHz RIN noise & \small{V1:SDB2\_B1s1\_PD1\_Blended} &  2019-04-05 & 2019-08-01 & 40-1000 \\
& & 2019-08-01 & 2019-08-01 & 40-2000 \\
& & 2019-08-01 & 2019-11-24 & 40-1000 \\
& & 2019-11-24 & 2020-03-10 & 40-2000 \\
& & 2020-03-10 & 2020-04-02 & 20-2000 \\
\hline \hline
Scattered light  &
\small{V1:SNEB\_B7\_DC} & 2019-04-05 & 2020-02-07 & 10-70 \\\cline{2-5}
noise at NE & \small{V1:SNEB\_B7\_DC\_D} & 2020-02-07 & 2020-04-02 & 10-70 \\
\hline \hline
Scattered light  & \small{V1:SWEB\_B8\_DC} & 2019-04-05 & 2020-02-07  & 10-70  \\\cline{2-5}
noise at WE &  \small{V1:SWEB\_B8\_DC\_D} &  2020-02-07 &  2020-04-02 & 10-70 \\
\hline
\end{tabular}}
\caption{\label{tab:noise_sub} Noises subtracted linearly during O3 within the $h(t)$ reconstruction process. Details are provided on the noise-witness channel considered to subtract each noise, the time period and the frequency band. Note that two channels, V1:SIB2\_B2\_8MHz\_I and V1:SIB2\_RFC\_PD2\_Audio, were setup to monitor and check the subtraction of the frequency noise during November 2019. The change of the noise witness channels used to subtract the scattered light in NE and WE is due to a change of the data type (float to double), done with the goal of reducing the numerical noise.
The blue italic line indicate the modification applied for the reprocessing of the end of O3a data.
}
\end{table}

\subsubsection{Performances during O3}

\begin{figure}[!ht]
	   \hspace{-0.5cm}   
	   \includegraphics[scale=0.40]{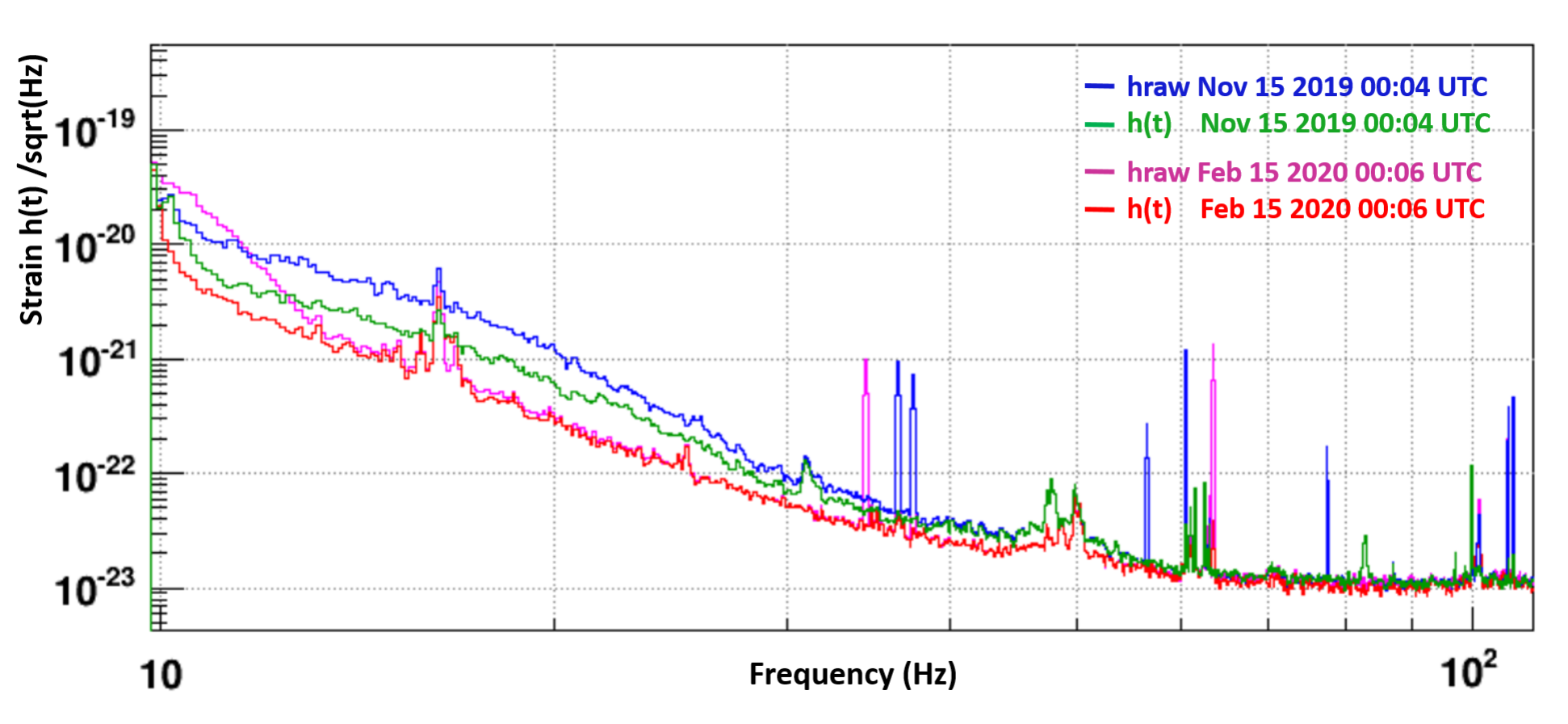} 
        \centering
    \caption{Spectra of $h(t)$ (green) and $h_{raw}(t)$ (blue) showing the main
    sensitivity improvement due to noise subtraction and calibration lines removal in the middle of O3\dv{. This improvement increased the BNS range by 7 Mpc}. Purple and red lines shows the same spectra near the end of O3, after some noise reduction has been done on the interferometer control loops.}
    \label{fig:noise_sub_psd}
\end{figure}

\begin{figure}[!ht]
        \centering
	    \includegraphics[scale=0.46]{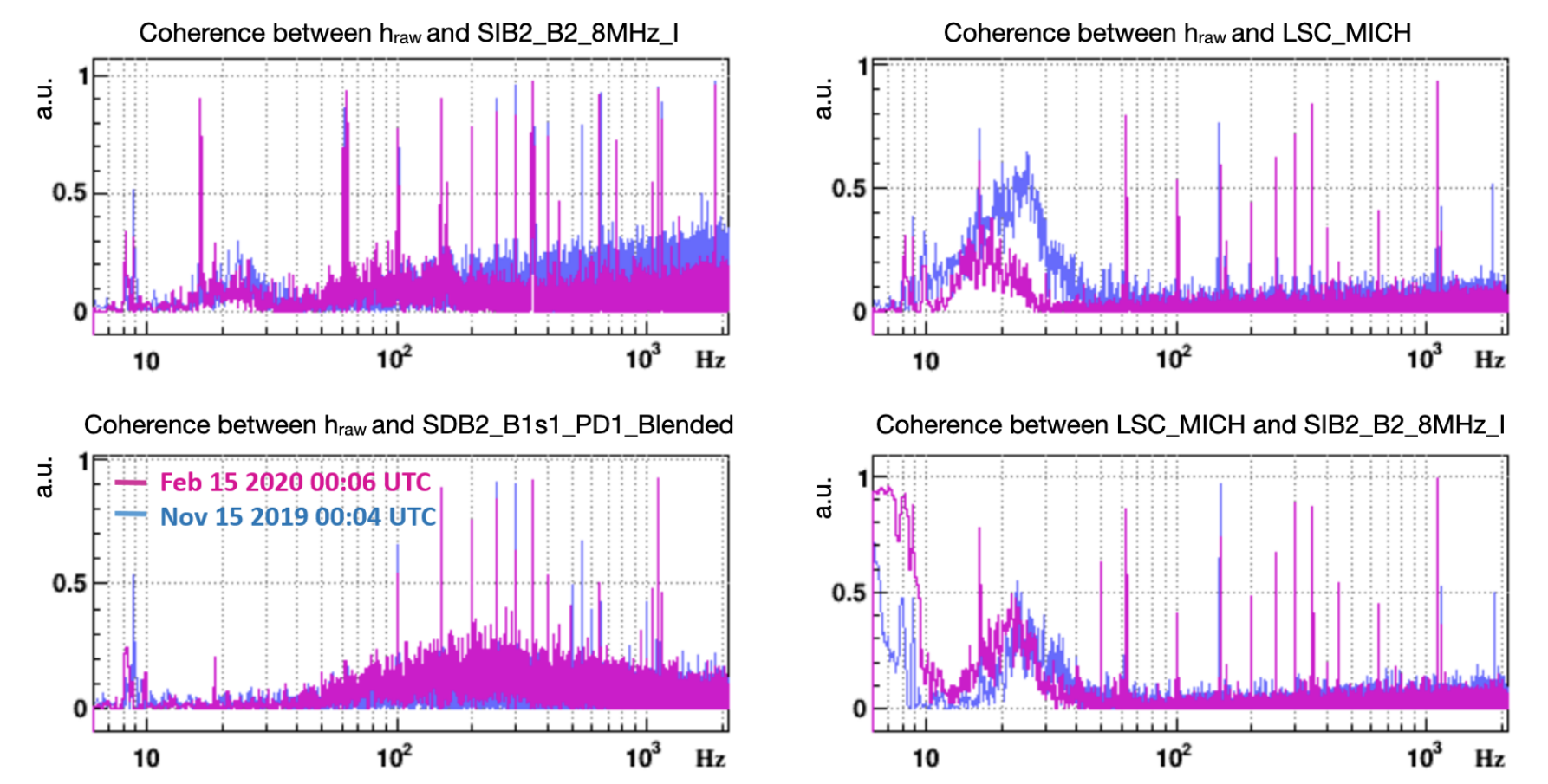}
    \caption{Coherence between $h_{raw}$(t) and the noise witness channels
    used in the middle of O3. The overlap between the frequency bands where LSC\_MICH
    and SIB2\_B2\_8MHz\_I are coherent with $h_{raw}$(t) required the use of the
    two frequency bands of Table \ref{tab:noise_sub}. Purple lines show the same coherence plots near the end of O3.}
    \label{fig:noise_sub_cohe}
\end{figure}

The use of these noise-witness channels in the noise subtraction of $h(t)$ during O3 allowed to improve the overall sensitivity, especially in the 10-50~Hz frequency band, \msa{as shown in figure~\ref{fig:noise_sub_psd}}, which translated into a gain of up to 7~Mpc on the binary neutron star (BNS) range. The main contributors to this improvement were the frequency noise (see figure~\ref{fig:noise_sub_cohe}) 
and the 56~MHz RIN noise which showed coherence with $h_{raw}$ over a large frequency band.

\begin{figure}[!htb]
  \begin{center}
    \subfigure[O3a and O3b best sensitivity curves.]{
      \includegraphics[width=0.8\linewidth]{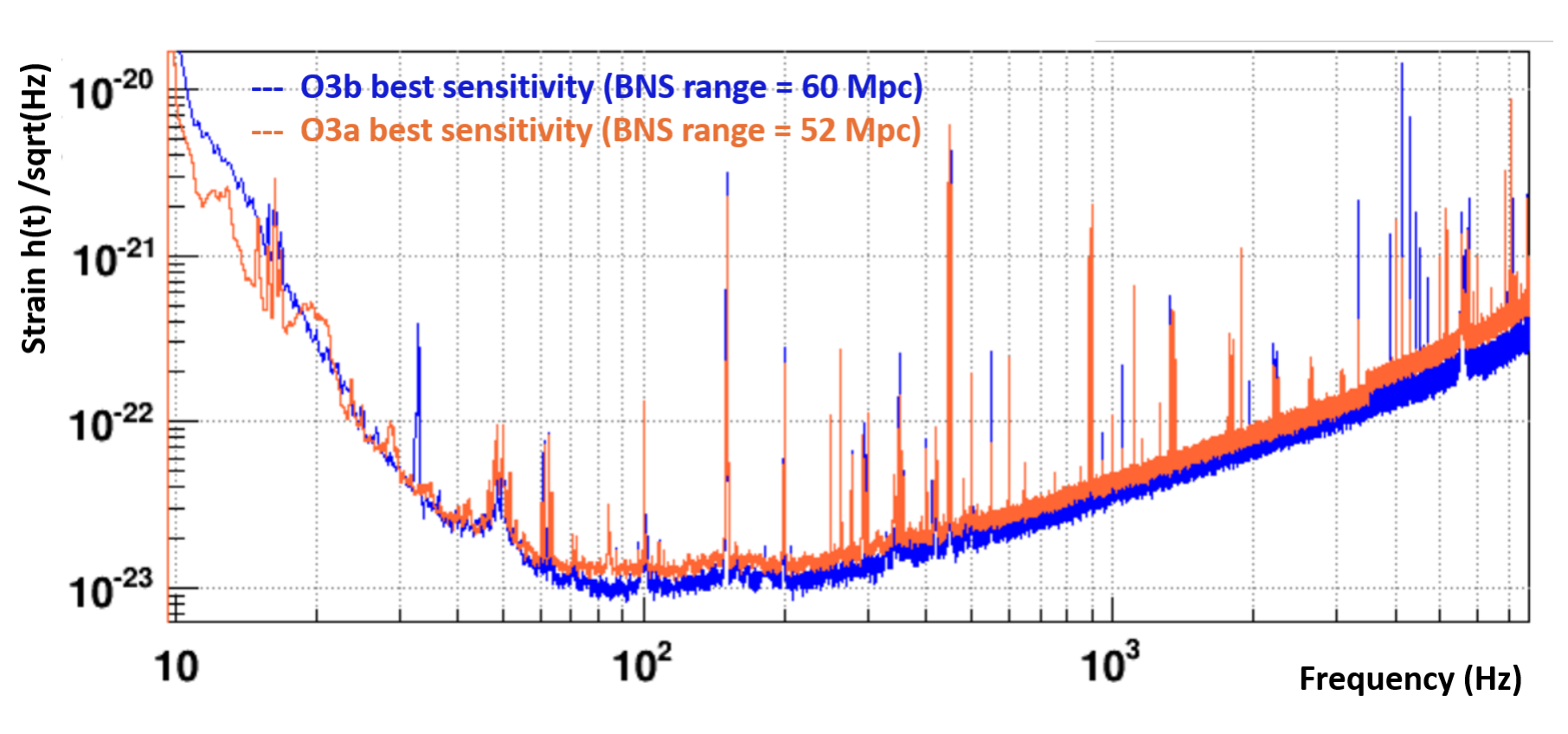} }
    \subfigure[BNS range evolutin during the O3 run.]{
      \includegraphics[width=0.8\linewidth]{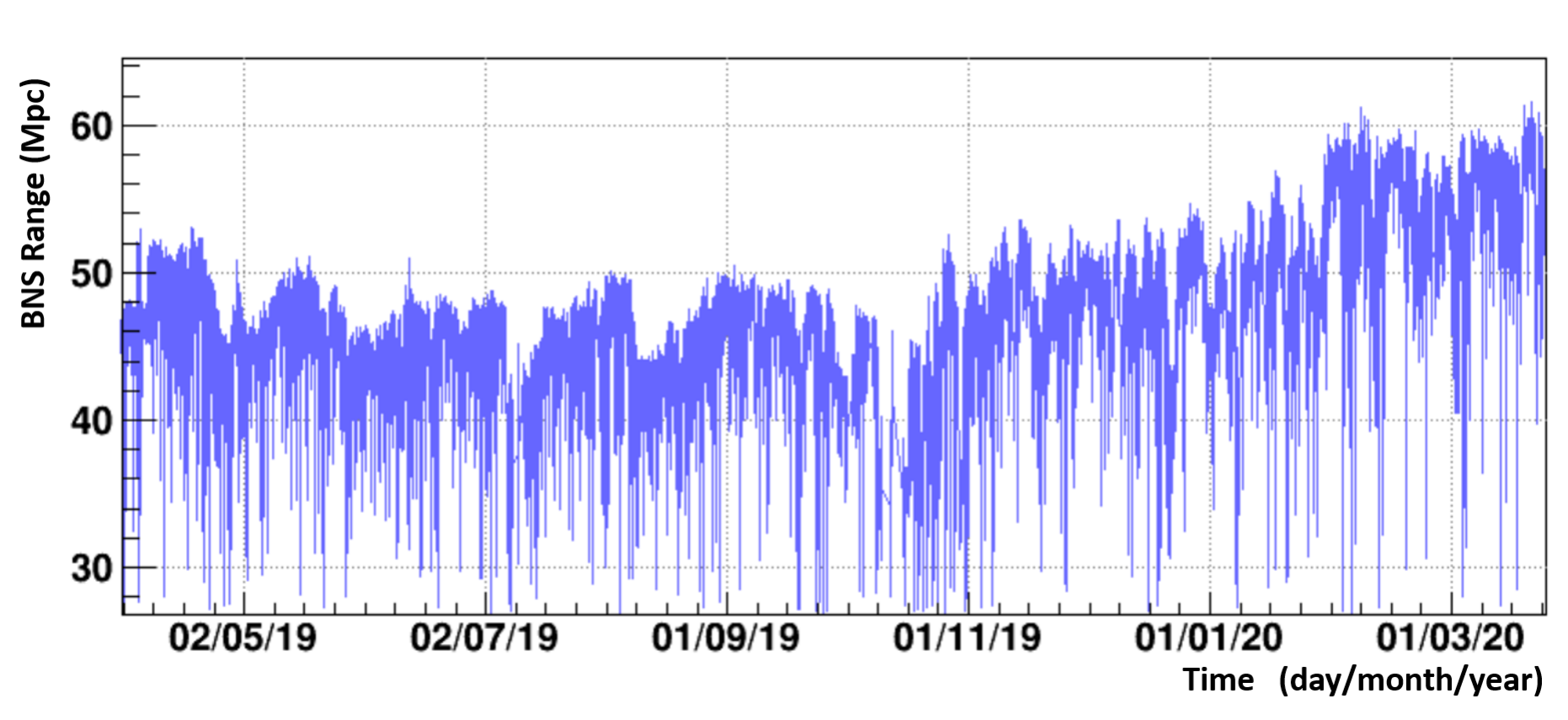} }
    \caption{Top: Best sensitivity curves (spectrum of h(t)) during O3a at GPS=1251317358 (orange) and O3b at GPS=1265155700 (blue). Bottom: Evolution of the BNS range over the O3 run. \dv{Short-duration and large amplitude fluctuations of the BNS range, down to 30 Mpc, are due to non-stationary noises. On a weekly time scale, BNS range fluctuations are mainly correlated with bad weather and seismic motion. The increasing trend over the O3b period is mainly due to detector's improvements and noises reduction done during the weekly commissioning periods.}
    }
    \label{fig:O3sensitivity}
  \end{center}
\end{figure}

Figure~\ref{fig:O3sensitivity} shows the sensitivity and the BNS range obtained during O3 thanks to the $h(t)$ reconstruction and after noise subtraction. \\

During the last two weeks of O3a, from 16 to 30 September 2019, commissioning activities that changed the interferometer working point multiple times were undertaken (tuning of etalon effect in the input cavity mirrors). As a consequence, the coupling of the some noise changed during that period, in particular for the frequency noise.
A reprocessing of the reconstruction was run offline for that period, with lower minimal frequency for this noise witness channel (see table~\ref{tab:noise_sub}). It resulted in a gain of up to 3~Mpc compared to the online noise subtraction~\cite{bib:calibnote_O3_Repro1A}.
The $h(t)$ time series associated to this reprocessing is called V1:Hrec\_hoft\_V1O3ARepro1A\_16384Hz.

\section{Estimation of h(t) uncertainties}
\label{sec:hrecerrors}

The Virgo $h(t)$ uncertainties during O3 were estimated using hardware injections and comparing the
reconstructed $h(t)$ to those known forced motion applied to the end mirrors. 
The excitation were applied on NE and WE mirrors via different actuators: electromagnetic actuators (channels $zHI_i$),
photon calibrators~\cite{bib:2020_PCalO3} (channels $\mathcal{P}^{ref}_i$) 
and Newtonian calibrators~\cite{bib:2020_NCalO3}. They consisted of sinusoidal signals at specific frequencies and of broad-band noise. 
In practice, the reconstructed channel $h_{clean}$ was used for this study,
since the hardware injections were subtracted in the final $h(t)$ channel.
But, for coherence with figures, we will write $h_{rec}(t)$ everywhere in this section.
By performing transfer functions from $h_{clean}(t)$ to $h(t)$, it was checked that
the subtraction of the hardware injections does not modify the results:
the transfer functions are equal to~1, except at the frequencies of the subtracted hardware injections (as expected).
Hence the estimated uncertainties are valid for both the intermediate $h_{clean}$ ($h_{rec}$) and the final $h(t)$ channel.\\

The $h(t)$ data stream is validated across the entire detection band, from 10~Hz to 1.5~kHz, at roughly weekly cadence. During this calibration process, the interferometer is in its standard working point but it is declared out of observing mode while excitation signals are applied to the end test masses. 
In addition, a few permanent sinusoidal excitation signals were applied with lower signal-to-noise ratio (order of~10) to monitor the stability of $h(t)$ uncertainties over the run. Such data provide a continuous measurement of the systematic error at only a few selected frequencies inside the most sensitive band of the detector (30~Hz to 400~Hz).
The overall O3 systematic uncertainties of the Virgo strain during O3 are estimated assessing the stability of the different measurements done every week (with high signal-to-noise ratio) at a larger number of frequencies and done permanently (with lower signal-to-noise ratio) at a few frequencies.

The $h(t)$ uncertainties analysis has been done separately for O3a and O3b.
The equivalent strain excitation $h_{inj}$ has been estimated from the raw excitation signals and the most up-to-date actuator models computed using all the O3 data as described in section~\ref{sec:calibActuators}.
Ideally, the transfer function from the excitation signal to the strain data 
is expected to be~1. Deviations from this ideal value may come from 
a bias in the $h(t)$ channel, 
but also from an inaccurate actuation model. 
Using different actuators and actuator models based on independent calibration techniques allows to conclude about the origin of the discrepancy to~1.\\

Some calibration variations of the PCal actuators have been seen during O3, mainly in relation with humidity~\cite{bib:2020_PCalO3} and have impacted the
TF from the PCal excitation to $h(t)$, as well as its temporal variations.
The electromagnetic actuators are expected to have a stable response.
Two different models have been computed for these actuators,
\lr{the reference one based on the PCal, and the one based on the free swinging Michelson technique used for cross validation}.\\

In the $h(t)$ reconstruction process, the optical model used for the cavity of finesse $\mathcal{F}$ is approximated by a simple cavity pole. 
The exact response is given in equation~28 of~\cite{bib:Rakhmanov_2008_shortwavelengthapprox}.
Up to 2~kHz, the approximation is good within 0.5\% in amplitude 
and it is biased by 13~\mus\ in timing~\cite{bib:2015_CalibApproximations}. 
When using the $h(t)$ channel later in the gravitational wave searches, 
another approximation is done by using the antenna response as the one of a point-like interferometer. 
The errors from both approximations nicely cancels out to within 0.1\% in amplitude and 3~\mus\ in timing below 2~kHz~\cite{bib:2015_CalibApproximations}. \\

Comparing the reconstructed detector strain, $h_{rec}$, directly to the injected mirror motion, $h_{inj}$ would reveal the error from the simple cavity pole approximation, mainly as a 10~\mus\ difference in the case of Virgo interferometer.
In this section, the estimated injected signal $h_{inj}$ has been
computed as
\begin{equation}
  h_{inj} = N \times \frac{A}{L_0} \times \e^{-2 j \pi f \tau } \times \frac{TF_{pole}}{TF_{true}}  
\end{equation}
with $N$ the noise excitation channel, $A$ the actuator transfer function, 
$L_0$ the arm cavity length and $\tau = 10$~\mus\ due to the difference between the interferometer response to a gravitational wave and to a real motion of the end mirror,
\lr{and $j^2 = -1$}.
$TF_{pole}$ is the simple pole approximation \lr{(with pole frequency $f_p$)} and $TF_{true}$ is the exact response defined as:
\begin{equation}
    TF_{pole} = \frac{1-j x}{1+x^2}\quad \textrm{with}\quad x = \frac{f}{f_p}
\end{equation}
\begin{equation}
    TF_{true} = \frac{1-\alpha}{1-\alpha\,\e^{-4 \pi f T}}\quad \textrm{with}\quad \alpha = \frac{\mathcal{F}}{\pi+\mathcal{F}}\quad \textrm{and}\quad T = \frac{L_0}{\text{c}}
\end{equation}
With this expression of $h_{inj}$, the ratio $h_{rec}/h_{inj}$ is expected to be~1, with no timing difference from the cavity response model approximation.

\subsection{Weekly measurements with sinusoidal excitation}
\label{sec:hrecerrors_weeklyO3a}

\begin{figure}[!ht]
    \centering
	\includegraphics[trim={0 0cm 0 0cm},clip,scale=0.42]{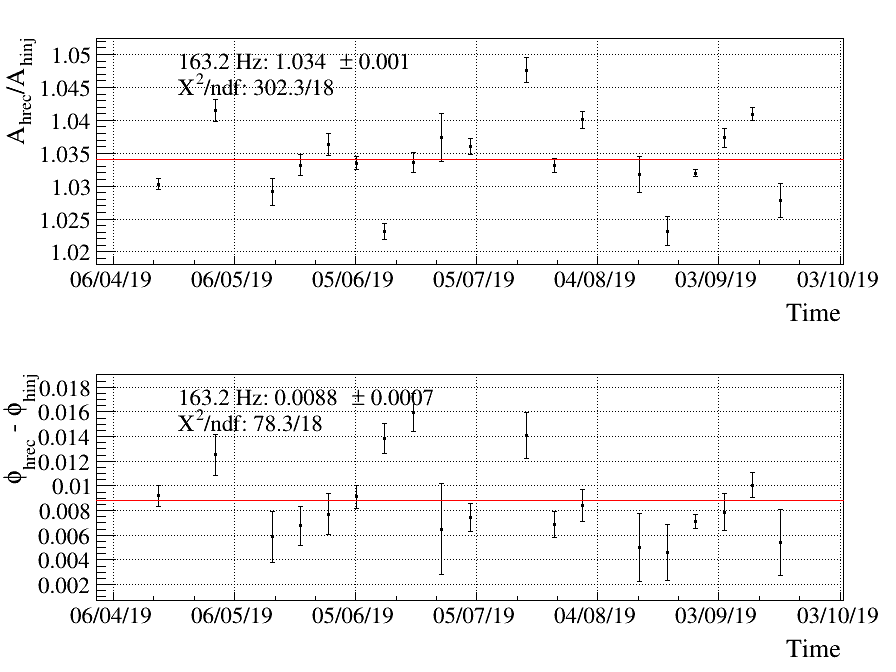} 
    \caption{Evolution of the $h_{rec}/h_{inj}$ transfer function as function of time at $163.2~$Hz during O3a, with $h_{inj}$ estimated from NE PCal.
    }
    \label{fig:hrechinj_NEpcal_vst}
\end{figure}

Injections of sinusoidal excitation on the end mirrors, between 10~Hz and 1.5~kHz, have been done every week and the transfer function from the injected equivalent strain $h_{inj}$ to the reconstructed strain $h_{rec}$ has been computed. 
Figure~\ref{fig:hrechinj_NEpcal_vst} shows the modulus and phase of this transfer function as function of time,  measured using PCal as actuator, at 163.2~Hz during O3a. \lr{When averaged over time, the modulus and phase are close to 1 and 0 as expected, but with 3\% and 8~mrad offset respectively: it gives the average bias on the reconstructed strain channel at this frequency. However, time variations are seen, which indicates the presence of additional and time-dependent errors on the strain channel.}\\

\begin{figure}[!ht]
    \begin{minipage}[t]{0.49\textwidth}
	   \includegraphics[width=0.99\linewidth]{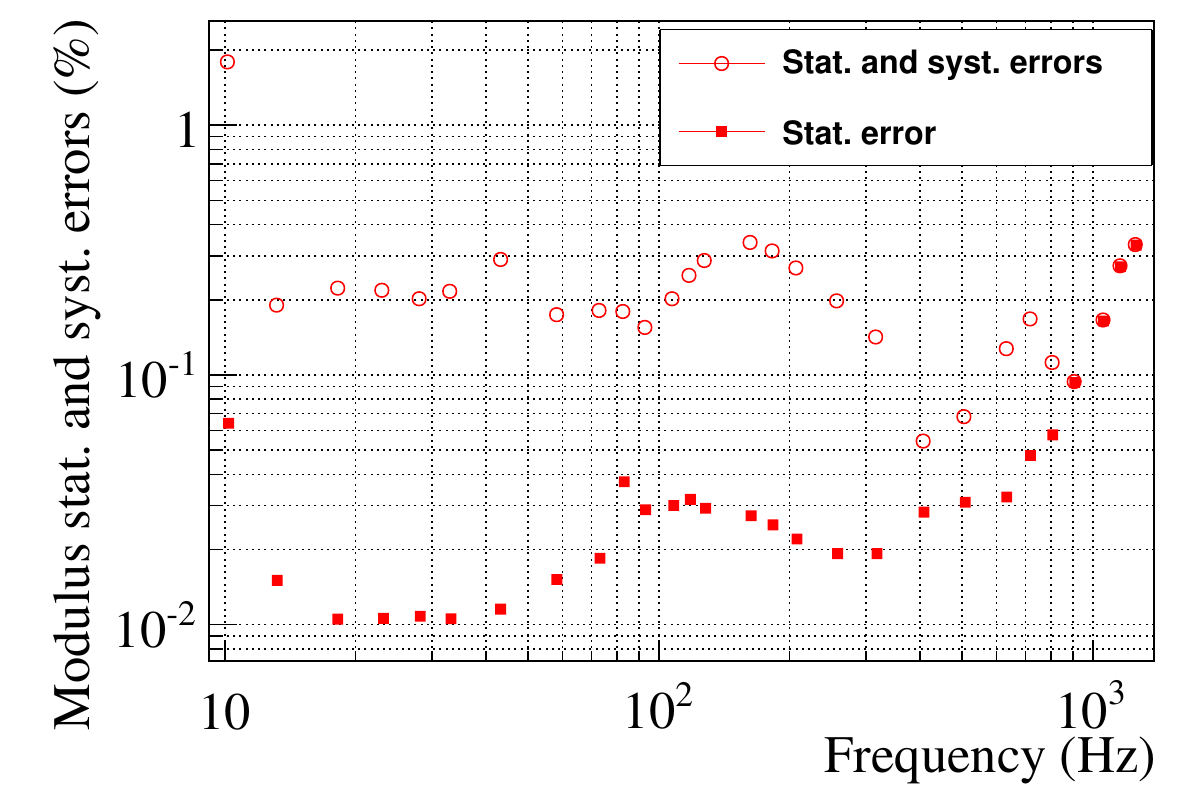} 
    \end{minipage}
    \begin{minipage}[t]{0.49\textwidth}
        \centering
	    \includegraphics[width=0.99\linewidth]{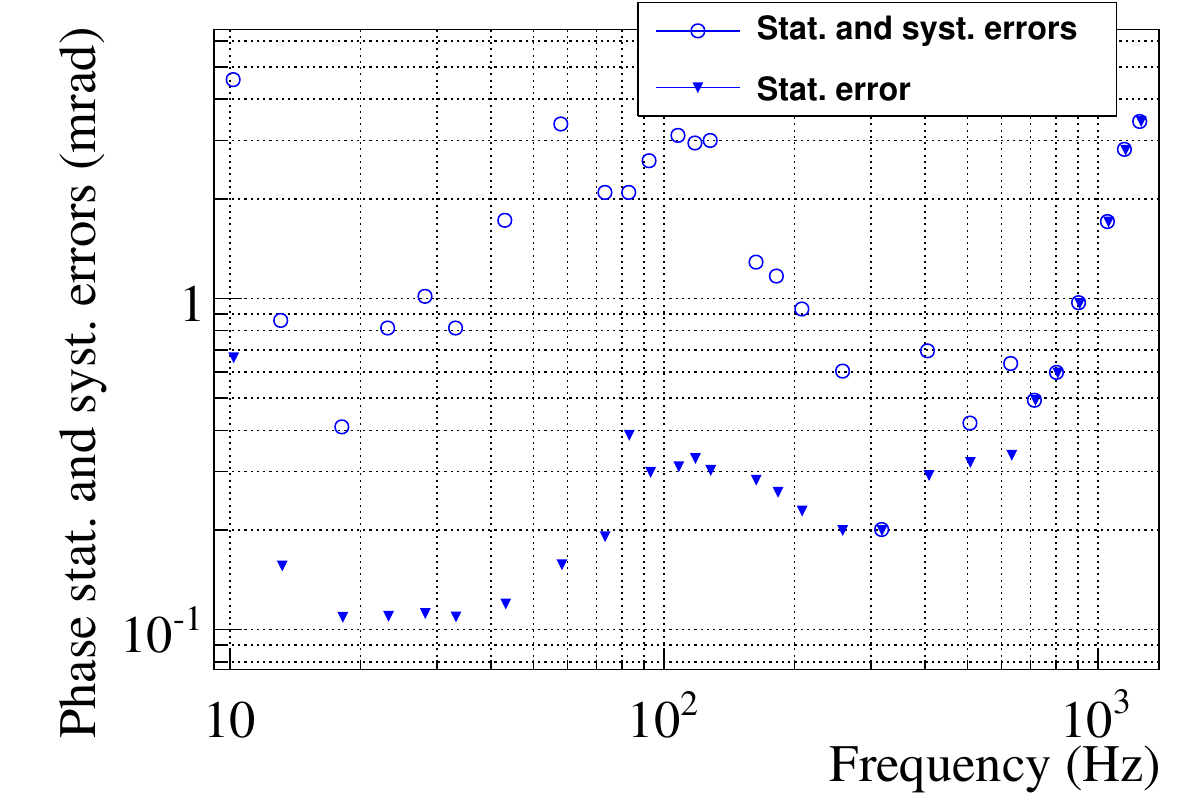}
    \end{minipage}
    \caption{Statistical and systematic errors estimation on the $h_{rec}/h_{inj}$ transfer function during O3a, with $h_{inj}$ estimated from NE PCal. The errors on the amplitude are expressed in [$\%$] and the errors on the phase are expressed in [mrad]. Left: Statistical errors on the amplitude (red filled squares) of the calibration transfer as a function of frequency and statistical plus systematic errors (red empty circles) estimation to have a p-value$~>0.05$. Right: Statistical errors on the phase (blue filled triangles) of the calibration transfer as a function of frequency and statistical plus systematic errors (blue empty circles) estimation to have a p-value$~>0.05$.}
    \label{fig:hrechinj_NEpcal_systO3a}
\end{figure}

The level of variations has been estimated following the same method as described in section~\ref{sec:TimeStabilitymirNE}. Figure~\ref{fig:hrechinj_NEpcal_systO3a} shows the estimated errors on the modulus and phase at the frequencies of the weekly sinusoidal injections.
Above 20~Hz, conservative systematic uncertainties
of 0.4\% on modulus and 4~mrad on phase can be used to take into account the differences between the weekly injections data.

Such analysis has been done for O3a and O3b separately, with hardware injections using the four type of actuators, PCal and electromagnetic actuators of NE and WE mirrors. 
The associated errors are summarized in the table~\ref{tab:hrechinj_systO3}.\\

Some sources of time variations have been identified.
Since they had small or no impact on the strain data channel, no reprocessing of the strain data has been done
and the variations have been included in the overall errors.
During O3a, the NE electromagnetic actuators had some digital communication issues from 10 to 20 June 2019:
it resulted in a phase for this actuator slightly modified during that period, by about 10~\mus.
However, the validation of $h(t)$ with the hardware injections performed with the other actuators has shown that it had small impact on the strain data channel bias, well lower than 1\% in amplitude and 1~mrad in phase.

During O3b, on 17 December 2019, the PCal digital anti-aliasing filter has been slightly modified, 
changing the PCal readout by 2.8~\mus. This had no impact on the strain data channel, but it had an impact
on the estimation of $h_{inj}$ when using the PCal for the analysis described in this section.

On 28 January 2020, in order to reduce their electronics noise, the two photodiodes used at the interferometer output and sensitive to the GW signal, as well as their analogue electronics, have been changed.
From then until the end of O3b, the reconstructed strain data channel had an extra 1~\mus\ delay. This is also well within the estimated timing uncertainty. 

\begin{table}[!ht]
\begin{center}
\begin{tabular}{|c|c|c|c|c|c|}
\hline 
 & Actuator             & NE PCal & NE EM & WE PCal & WE EM \\ 
\hline
O3a & Modulus error      & 0.4\%   & 0.3\%   & 0.4\%  & 0.3\%  \\
O3a & Phase error (mrad) & 4       & 3       & 4      & 3      \\
\hline
O3b & Modulus error      & 0.4\%   & 0.6\%   & 0.5\%  & 0.5\%  \\
O3b & Phase error (mrad) & 4       &   5     & 6      & 6      \\
\hline
\end{tabular}
\caption{\label{tab:hrechinj_systO3} 
\lr{Summary of the uncertainties on the $h(t)$ bias, estimated from time variations of weekly measurements of $h_{rec}/h_{inj}$ transfer function during O3a and O3b, 
in the range 20~Hz to 1500~Hz.
Four different actuators have been used to excite the NE and WE mirrors and to estimate $h_{inj}$: 
for both mirrors, the photon calibrator (PCal) actuator and the electromagnetic (EM) actuator have been used.}
Similar type of variations have been estimated with permanent monitoring 
as described later in section~\ref{sec:hrecerrors_permmoni}.
}
\end{center}

\end{table}

\begin{figure}[!ht]
    \centering
	\includegraphics[trim={0 0cm 0 0cm},clip,scale=0.52]{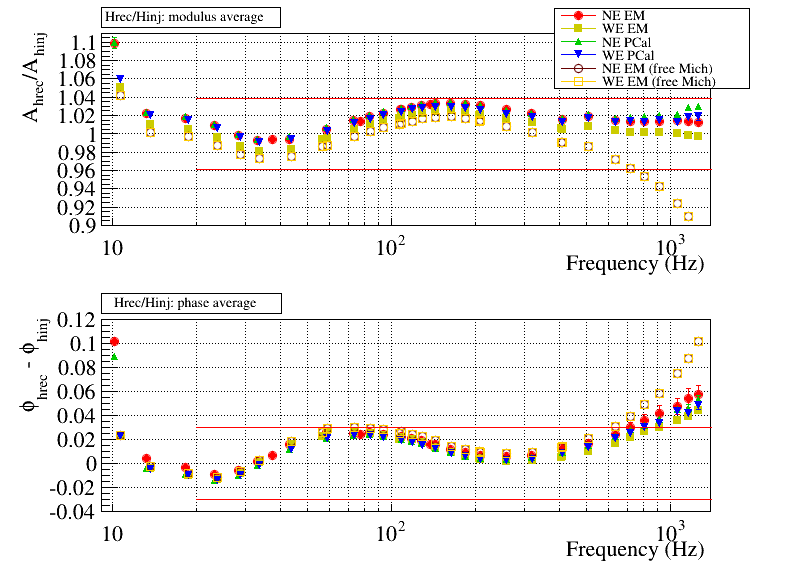}
    \caption{\label{fig:hrechinj_all_O3a}
    Average of O3a weekly $h_{rec}/h_{inj}$ transfer functions estimated from hardware injections using four different actuators: electromagnetic (EM) and photon calibrator (PCal) for both mirrors. Two different calibration models of electromagnetic actuators have been used: the reference one, based on PCal technique, and an independent one, based on free swinging Michelson technique.
    Only statistical errors are shown.
    The red lines indicate bands of $\pm3.9\%$ in modulus and $\pm30~\text{mrad}$ in phase.
    }
\end{figure}

\begin{figure}[!ht]
    \centering
	\includegraphics[trim={0 0cm 0 0cm},clip,scale=0.52]{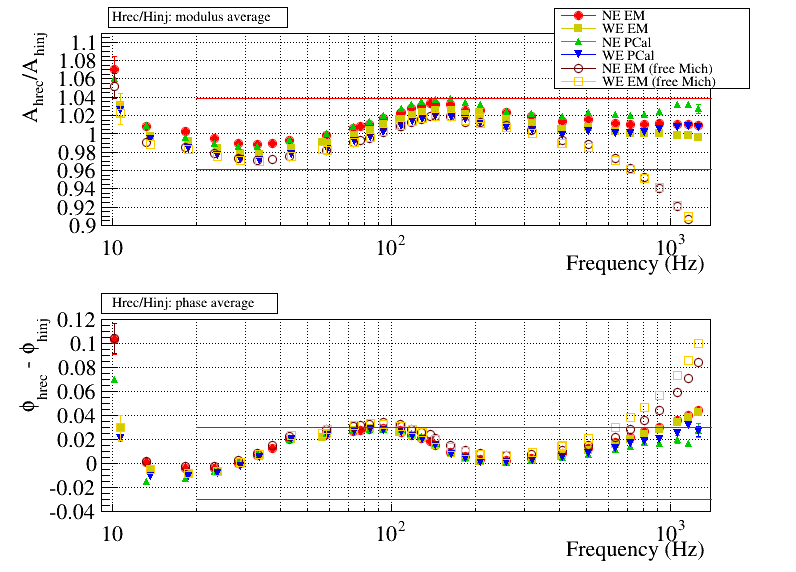}
    \caption{Average of O3b weekly $h_{rec}/h_{inj}$ transfer functions estimated from hardware injections using four different actuators: electromagnetic (EM) and photon calibrator (PCal) for both mirrors. Two different calibration models of electromagnetic actuators have been used: the reference one, based on PCal technique, and an independent one, based on free swinging Michelson technique.
    Only statistical errors are shown.
    The red lines indicate bands of $\pm3.9\%$ in modulus and $\pm30~\text{mrad}$ in phase.
    }
    \label{fig:hrechinj_all_O3b}
\end{figure}

The errors being small, all the weekly measured transfer functions have been averaged over O3a and over O3b
and are shown in figures~\ref{fig:hrechinj_all_O3a} and~\ref{fig:hrechinj_all_O3b} respectively.
The results obtained for the four different actuators are included. For the electromagnetic (EM) actuators,
two different models have been used: the reference one based on the PCal technique,
and the one extracted from the free swinging Michelson technique.

A systematic frequency-dependent bias is seen, up to 1.5~kHz, using the four different actuators and using independent calibration based on both photon calibrator and free swinging Michelson techniques.  
It has been seen also up to 120~Hz with a third independent actuator, the Newtonian calibrator~\cite{bib:2020_NCalO3}.
It is thus excluded that the bias comes from errors on the estimated injected signals $h_{inj}$,
and it hence confirms that there is a true bias in the reconstructed $h(t)$ strain channel.
The red lines on figures~\ref{fig:hrechinj_all_O3a} and~\ref{fig:hrechinj_all_O3b} show the (symmetric)
bands that contain all the PCal-based measurements, within $\pm3.9\%$ in modulus and $\pm30~\text{mrad}$ in phase.
At high frequency, the phase goes out of this band as it can be interpreted as a bias on the timing of $h(t)$: \lr{for a mistiming of $\Delta \tau$, the phase bias increases with frequency $f$ as $\Delta \Phi = -2\pi \times f \times \Delta\tau$.}
The reconstructed $h(t)$ channel looks to be slightly in advance, by about 7~\mus\ and 6~\mus\ during O3a and O3b respectively. 

The actuator calibration based on free swinging Michelson technique is less sensitivity at high frequency and
the fitted model is less precise, which explains the increasing deviation from the measurements based on photon calibration 
(see section~\ref{sec:calib_michelson_compare}). 
The actuation delay is also less constrained using the free swinging Michelson technique: 
a difference of about $-5$~\mus\ is seen with respect to the photon calibration results. 
The larger phase bias seen on the measurements based on free swinging Michelson technique mainly
comes from this extra-delay, which corresponds to about 3~mrad around the maximum of the bias at 100~Hz.

Another method to cross-check the absolute amplitude of the reconstructed $h(t)$ strain
using light scattered by the optical benches in transmission of the NE and WE mirrors
has been proposed in~\cite{bib:was2021end}. A first estimate, but still with~5\% systematic uncertainties, confirmed the reconstructed $h(t)$ amplitude up to 30~Hz.\\

\lr{
At high frequency, in particular above 1.5~kHz, the $h(t)$ reconstruction
is much simpler since the interferometer controls are negligible in this band:
the reconstruction is dominated by the output power $\mathcal{P}_{DC}$ signal
and the models for the photodiode readout and interferometer optical response.
Hence, no unexpected additional bias is foreseen in particular in the 1.5~kHz to 2~kHz band of interest for high frequency data analysis.
The fact that, in the range 500~Hz to 1.5~kHz, the measured $h_{rec}/h_{inj}$ ratio is mainly flat in modulus and following a delay in phase confirms that this assumption seems correct.
In addition, injections at higher frequencies were done with the PCal, in the range 4 to 7~kHz. In this range, the PCal actuation response is known to within about 15\% only. They also confirm that the amplitude of the reconstructed detector strain is still under control within uncertainties lower than 15\% at 4~kHz~\cite{bib:estevez:tel-03337258}.
From these information, we can extend up to 2~kHz, the $h(t)$ uncertainties measured up to 1.5~kHz.
}

\subsection{Weekly measurements with broadband excitation}
\label{sec:hrecerrors_weekly_broadband}
Injections of broadband signal on the end test masses have been run every week and the transfer functions from the injected equivalent strain $h_{inj}$ to the reconstructed strain $h_{rec}$ have been computed too. They do not allow a monitoring of time variation as precise as with high signal to noise ratio sinusoidal signals, as described in previous section. The comparison of weekly measurements is thus limited by statistical errors. But they allow to estimate potential strain data errors in the whole detection frequency band. No particular issue has been found except around 50~Hz
as shown below. 
Two different errors on $h(t)$ have been found, \lr{both related to control loops using the main output of the interferometer, $\mathcal{P}_{DC}$, as error signal:}
\begin{enumerate}
    \item one at 50~Hz, related to the mains power line and the control loop set up during~O3 for its subtraction,
    \item one in the band 46-51~Hz related to the online control loop set up between O3a and O3b to damp some mechanical resonances of the suspensions at 48~Hz.
\end{enumerate}
It was first seen analysing the Newtonian calibrator (NCal)
O3b data~\cite{bib:2020_NCalO3} and was then confirmed with further analysis of injections done with PCal and EM actuators
as described in this section.\\

\begin{figure}[bph!]
  \begin{center}
    \subfigure[O3a, injections with NE EM actuator]{
      \label{fig:CheckHrec_MirNE_O3a}
      \includegraphics[angle=0,width=0.48\linewidth]{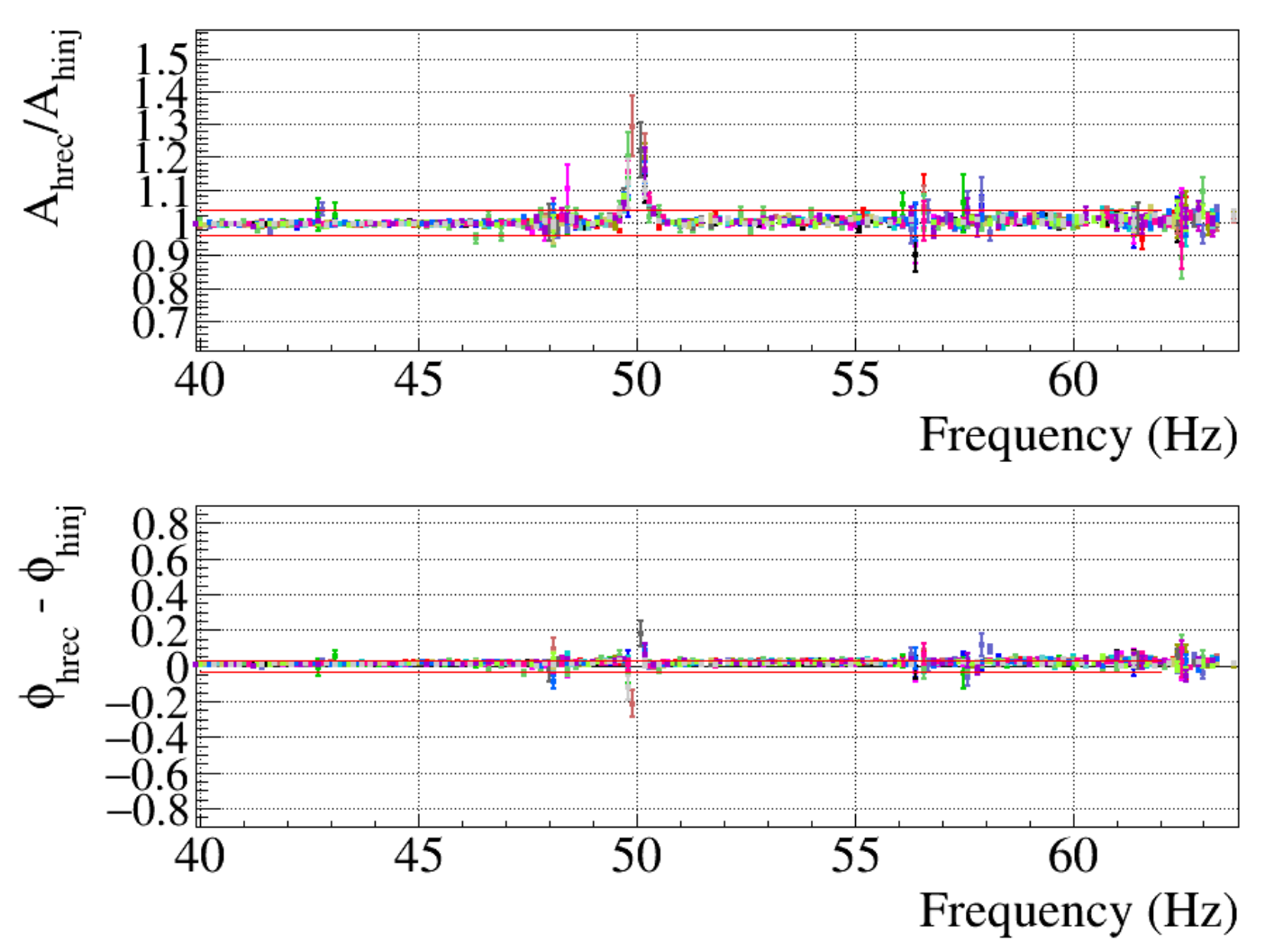}}
    \subfigure[O3b, injections with NE EM actuator]{
      \label{fig:CheckHrec_MirNE_O3b}
      \includegraphics[angle=0,width=0.48\linewidth]{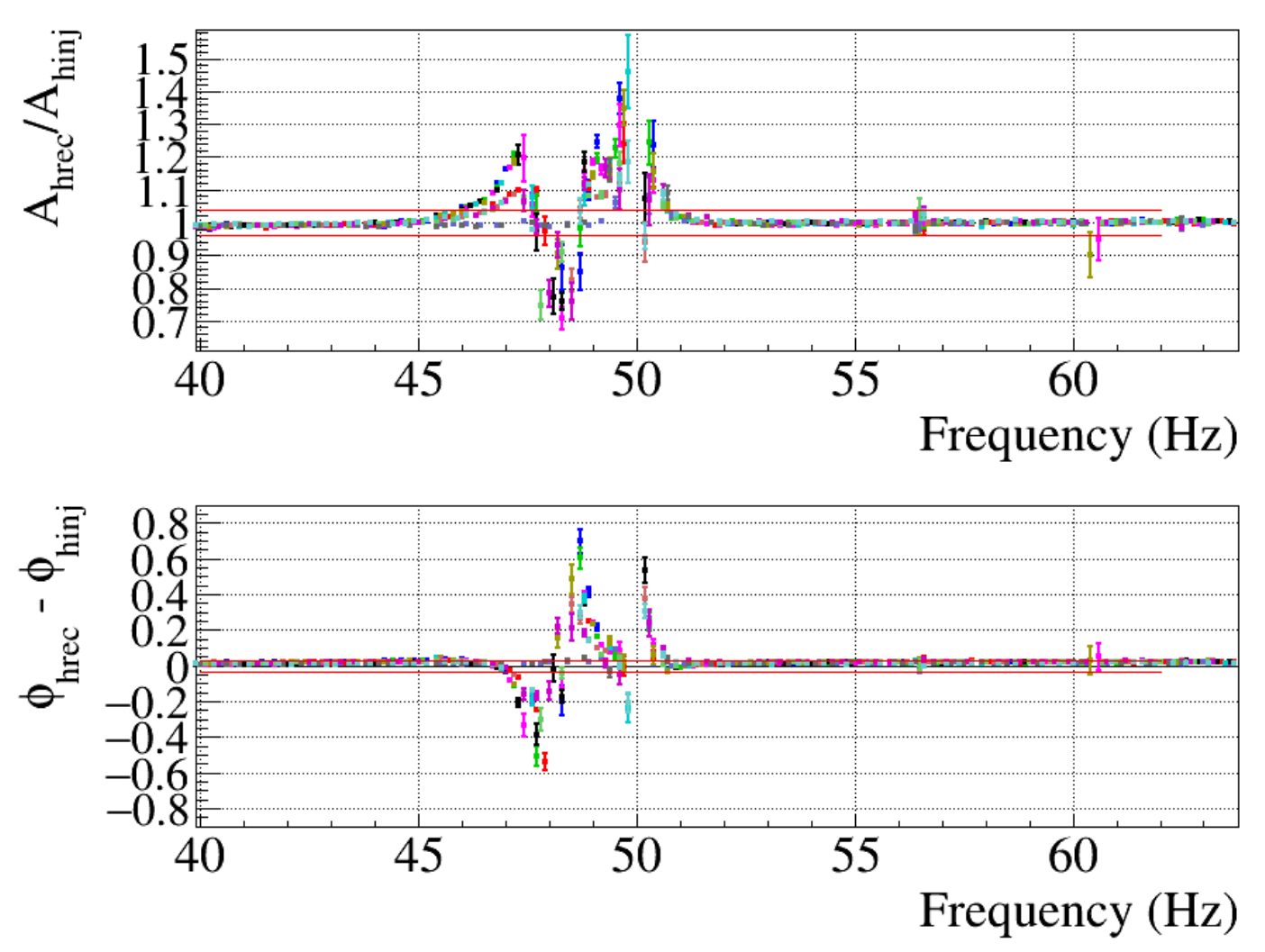}}\\
    \subfigure[O3a, injections with NE EM actuator: zoom]{
     \label{fig:CheckHrec_MirNE_O3a_Zoom}
      \includegraphics[angle=0,width=0.48\linewidth]{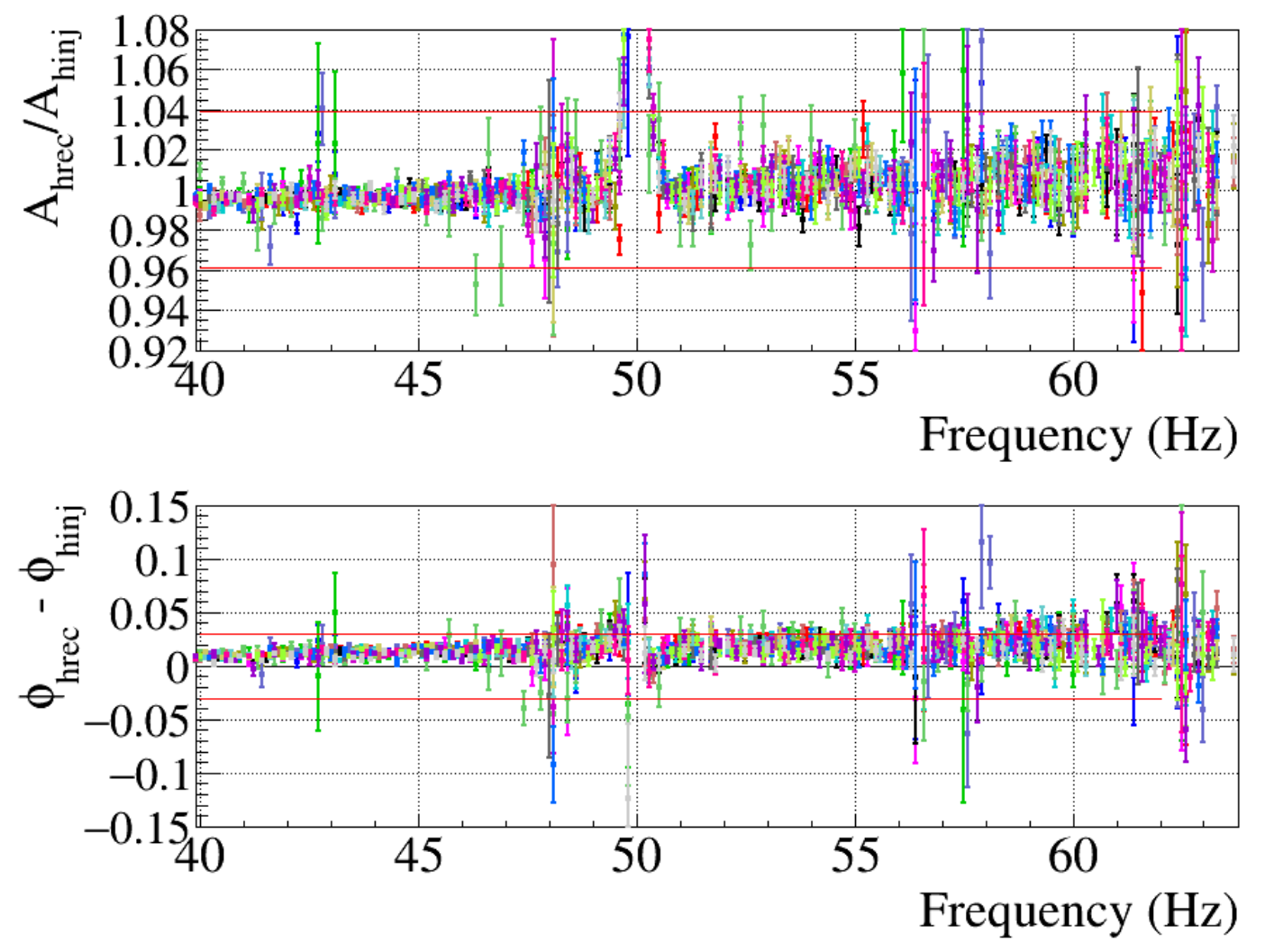}}
    \subfigure[O3b, injections with NE EM actuator: zoom]{
      \label{fig:CheckHrec_MirNE_O3b_Zoom}
      \includegraphics[angle=0,width=0.48\linewidth]{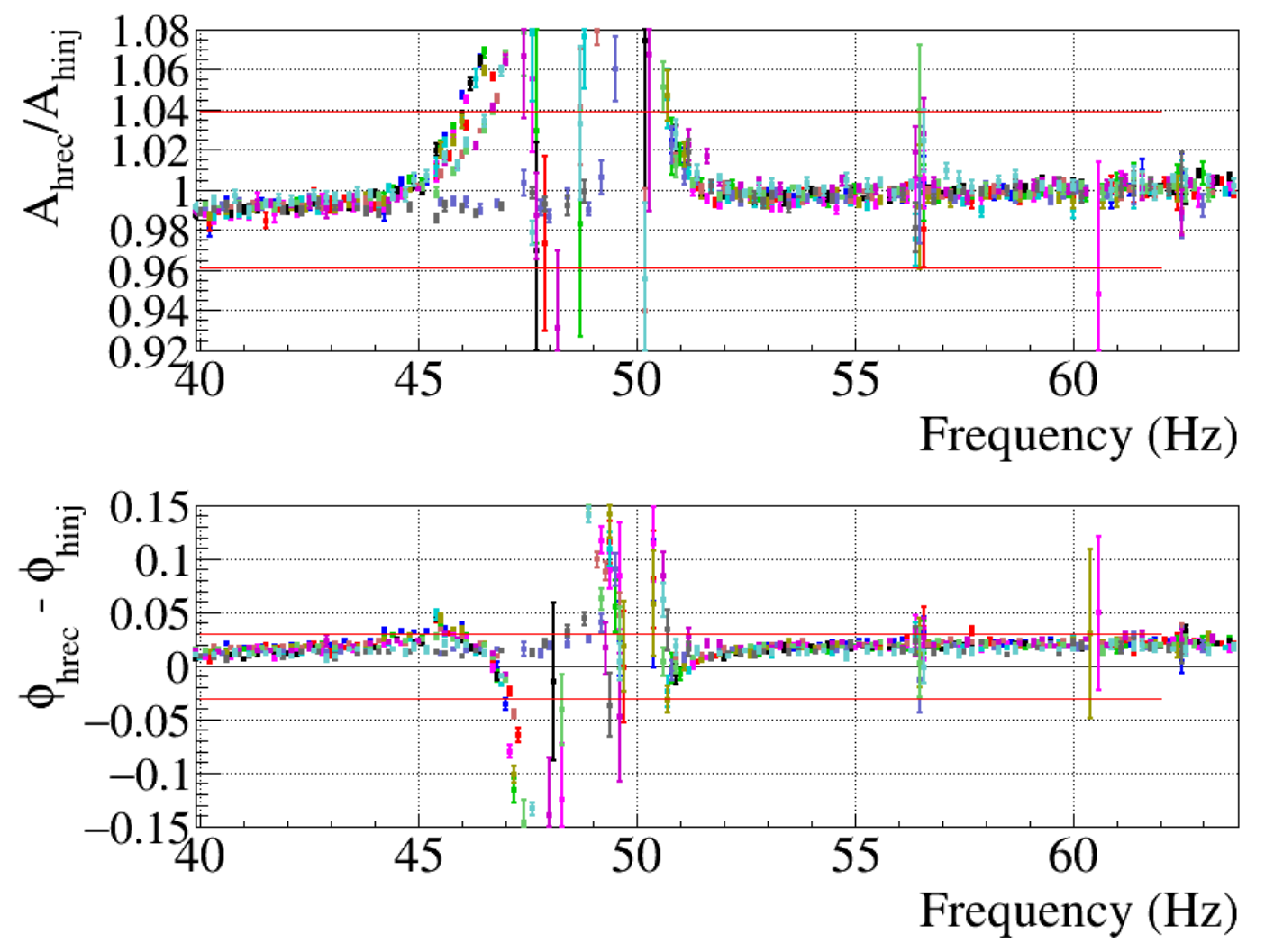}}\\
    \caption{Weekly measurements of $h_{raw}/h_{inj}$ transfer function using the NE EM actuator, shown in different colors,
      zoomed in the band 40 to 60~Hz.
      Left: O3a. Right: O3b.
      Top: full y-scale.
      Bottom: zoom on y-scale.
      Red lines indicate bands of $\pm3.9\%$ in modulus and $\pm30~\text{mrad}$ in phase.
      }
    \label{fig:CheckHrec_MIR_NE_50Hz}
  \end{center}
\end{figure}

Figure~\ref{fig:CheckHrec_MIR_NE_50Hz} shows the ratio $h_{rec}/h_{inj}$ for hardware injections of broadband noise 
done with the NE mirror electromagnetic actuator, during O3a (left) and O3b (right). 
Data from different weeks are shown in different colors.
The results are similar when using other end mirror actuators.
On figures~\ref{fig:CheckHrec_MirNE_O3a} and~\ref{fig:CheckHrec_MirNE_O3a_Zoom}, 
one can see that, during O3a, only a few bins around 50~Hz are not within the margin shown as red lines 
($\pm3.9\%$ in modulus and $\pm30~\text{mrad}$ in phase), mainly due to a lack of coherence between $h_{rec}$ and $h_{inj}$.

During O3b, an additional effect appeared, with a bias in modulus and phase on a larger frequency band, 
which reaches 40\% in modulus and 600~mrad in phase as shown in figure~\ref{fig:CheckHrec_MirNE_O3b}. 
From the zoom shown in figure~\ref{fig:CheckHrec_MirNE_O3b_Zoom}, one can see that the bias exceeds the standard limits between 46~Hz and 51~Hz.

These effects have been taken into account when estimating the strain data uncertainty in section~\ref{sec:hrecerrors_conclusions}.

\newpage

\subsection{Permanent monitoring of h(t) during O3}
\label{sec:hrecerrors_permmoni}

During O3, a set of 12 permanent sinusoidal signals, listed in tables~\ref{tab:HIlines_O3a}~and~\ref{tab:HIlines_O3b},
were injected on the end mirrors via the PCal and electromagnetic actuators. They allow a continuous comparison of the reconstructed strain data $h_{rec}$ with the injected equivalent strain $h_{inj}$ in the most sensitive band of the detector, between 35~Hz and 400~Hz. Some of those sinusoidal injections are in the regions where the bias of the strain channel is the highest (100 to 200~Hz for the amplitude and 60 to 90~Hz for the phase). Hence they are used to monitor any time variations of this bias.

\begin{figure}[!ht]
  \centering
   \includegraphics[trim={0 0cm 0 0cm},clip,scale=0.6]{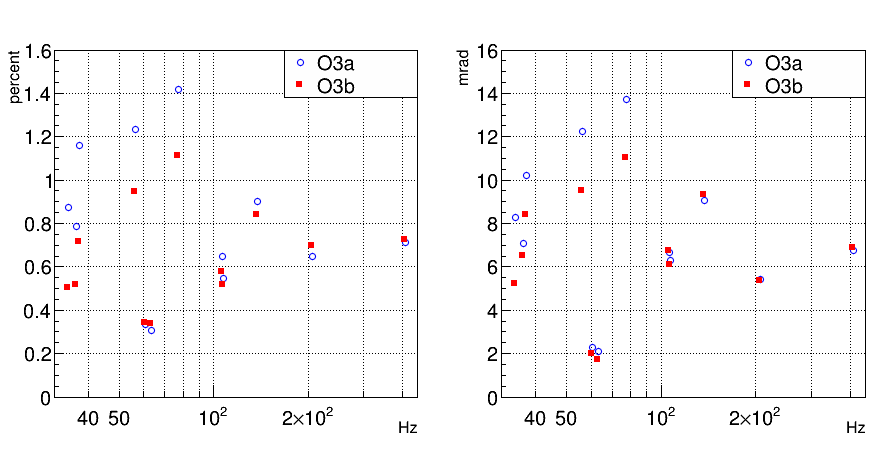}\\
    \caption{Left: systematic uncertainties on hrec/hinj module computed as the quadratic difference between the measured uncertainties and the computed statistical uncertainties. Statistical uncertainties have been computed using a SNR estimated at GPS=1251525400 for~O3a (blue empty circles) and GPS=1265000000 for~O3b (red filled squares)~\cite{bib:2021_TFMoni_O3}. 
    It was checked that the SNR during the run was consistent with these estimations within 20\%.
    Right: systematic uncertainties on hrec/hinj phase computed as the quadratic difference between the measured uncertainties and the computed statistical uncertainties. Statistical uncertainties have been computed on the basis of~\cite{2011CQGra..28b5005A}. 
    }
    \label{fig:hrechinj_systematics}
\end{figure}

The time evolution of this $h_{rec}$ to $h_{inj}$ comparison provides an other estimate of the error on $h(t)$.
The modulus and phase of the $h_{rec}/h_{inj}$ transfer functions at the twelve injected frequencies
were computed online using a moving average of twelve 10~s long FFTs and provided in the online data stream as channels sampled at 1~Hz~\cite{bib:TFMoni_doc,bib:2021_TFMoni_O3}. 
The $h_{inj}$ signals were estimated with online models set at the beginning of~O3 and were not updated in this processing during O3. As such, the monitoring of these channels is a test of the possible variations of the reconstructed strain data,
as well as of the small variations of the actuator responses as described earlier.
Yet, this method does not enable to check the absolute value of the strain data bias: they are not precise witness of the absolute value of the strain data bias.
The distributions of the online modulus and phase have been built over O3a and O3b
and their standard deviations are reported in tables~\ref{tab:HIlines_O3a}~and~\ref{tab:HIlines_O3b}.
For every injected signal, the typical signal-to-noise ratio during O3a is also given, as well as the expected standard deviation of the modulus in case of statistical fluctuations only. 
The measured variations during O3a are all larger than the expected statistics fluctuations only: 
this indicates that these monitoring channels indeed highlight time variations of the $h_{rec}/h_{inj}$ ratio.
These variations can be considered as the systematic uncertainties on this ratio and have been estimated by subtracting quadratically the expected statistical fluctuations from the measured standard deviation. 
They are shown, for O3a and O3b, in the plots of figure~\ref{fig:hrechinj_systematics}.
For~O3a and~O3b, the modulus systematic uncertainty at 137.5~Hz is estimated to be 0.9\%,
while the phase systematic uncertainty is estimated to be 9~mrad.
The maximum variations, at the level of 1.4\% in modulus and 14~mrad phase during~O3a, include contributions from both $h(t)$ and the actuator responses. They have been used to estimate conservative uncertainties on~$h(t)$ as described in the following section.

\newpage

\begin{table}[!ht]
\centering
\begin{tabular}{|l|l|l|l|l|l|l|l|}
\hline 
Actuator  &  Line freq.  &  Line SNR  &  $\sigma_{stat}$ & O3a $\sigma_{mod}$ &  O3a $\sigma_{phi}$ \\
\hline \hline
NE EM    & 37.5 Hz   &  3    & 1.06\%  & 1.57\%  & 15.8 mrad   \\
NE EM    & 77.5 Hz   &  4.5  & 0.70\%  & 1.58\%  & 15.5 mrad   \\
NE EM    & 107.5 Hz  &  13   & 0.25\%  & 0.60\%  & 6.8 mrad    \\
NE EM    & 137.5 Hz  &  7.5  & 0.43\%  & 1.00\%  & 10.1 mrad   \\
\hline
NE PCal  & 34.5 Hz   &  6    & 0.37\%  & 0.95\%  & 9.1 mrad    \\
NE PCal  & 63.5 Hz   &  25   & 0.09\%  & 0.32\%  & 2.3 mrad    \\
\hline
WE EM    & 56.5 Hz   &  5    & 0.66\%  & 1.40\%  & 13.9 mrad   \\
WE EM    & 106.5 Hz  &  11   & 0.29\%  & 0.71\%  & 7.3 mrad    \\
WE EM    & 206.5 Hz  &  11   & 0.29\%  & 0.71\%  & 6.2 mrad    \\
WE EM    & 406.5 Hz  &  9.5  & 0.34\%  & 0.79\%  & 7.6 mrad    \\
\hline
WE PCal  & 36.5 Hz   &  5    & 0.44\%  & 0.90\%  & 8.4 mrad    \\
WE PCal  & 60.5 Hz   &  22   & 0.10\%  & 0.35\%  & 2.5 mrad    \\
\hline
\end{tabular}
\caption{\label{tab:HIlines_O3a} Sinusoidal permanent hardware injections used during~O3a to monitor the accuracy of $h(t)$ reconstruction. 
For the four different end test mass actuators, the injected line frequency is given 
with the typical signal-to-noise ratio (SNR) estimated during O3a, 
the expected standard deviation $\sigma_{stat}$ of the $h_{rec}/h_{inj}$ modulus in case
of statistical fluctuations only, 
and the measured standard deviations
of the modulus ($\sigma_{mod}$) and of the phase ($\sigma_{phi}$) of $h_{rec}/h_{inj}$ 
during O3a.}
\end{table}

\begin{table}[!ht]
\centering
\begin{tabular}{|l|l|l|l|l|l|l|l|}
\hline 
Actuator  &  Line freq.  &  Line SNR  &  $\sigma_{stat}$ & O3b $\sigma_{mod}$  & O3b $\sigma_{phi}$ \\
\hline \hline
NE EM    & 37.5 Hz   &  11  & 0.29\%  & 0.77\%  & 8.9 mrad   \\
NE EM    & 77.5 Hz   &  7   & 0.46\%  & 1.20\%  & 12.0 mrad  \\
NE EM    & 107.5 Hz  &  16  & 0.20\%  & 0.55\%  & 6.4 mrad   \\
NE EM    & 137.5 Hz  &  9   & 0.36\%  & 0.91\%  & 10.0 mrad  \\
\hline
NE PCal  & 34.5 Hz   &  12  & 0.18\%  & 0.53\%  & 5.5 mrad   \\
NE PCal  & 63.5 Hz   &  41  & 0.06\%  & 0.34\%  & 1.8 mrad   \\
\hline
WE EM    & 56.5 Hz   &  7   & 0.46\%  & 1.05\%  & 10.6 mrad  \\
WE EM    & 106.5 Hz  &  14  & 0.23\%  & 0.62\%  & 7.1 mrad   \\
WE EM    & 206.5 Hz  &  15  & 0.22\%  & 0.73\%  & 5.8 mrad   \\
WE EM    & 406.5 Hz  &  12  & 0.27\%  & 0.77\%  & 7.4 mrad   \\
\hline
WE PCal  & 36.5 Hz   &  14  & 0.16\%  & 0.54\%  & 6.7 mrad   \\
WE PCal  & 60.5 Hz   &  47  & 0.04\%  & 0.34\%  & 2.0 mrad   \\
\hline
\end{tabular}
\caption{\label{tab:HIlines_O3b} Sinusoidal permanent hardware injections used during~O3b to monitor the accuracy of $h(t)$ reconstruction. 
    \lr{For the four different end test mass actuators, the injected line frequency is given 
    with the typical signal-to-noise ratio (SNR) estimated during O3b, 
    the expected standard deviation $\sigma_{stat}$ of the $h_{rec}/h_{inj}$ modulus in case of statistical fluctuations only,
    and the measured standard deviations 
    of the modulus ($\sigma_{mod}$) and of the phase ($\sigma_{phi}$) of $h_{rec}/h_{inj}$ during O3b.}
    }
\end{table}

\subsection{h(t) uncertainties}
\label{sec:hrecerrors_conclusions}

The uncertainties on the reconstructed strain channel $h(t)$ can be
estimated from the results given in the previous sections.
For O3, they have been estimated as frequency-independent symmetric uncertainties in the band 20~Hz to 2~kHz,
except for a small band around 50~Hz where they are significantly larger.\\

First, a systematic frequency-dependent bias is present in the reconstructed channel $h(t)$,
as shown in figures~\ref{fig:hrechinj_all_O3a} and~\ref{fig:hrechinj_all_O3b}:
it is observed, regardless of the actuator used for the measurement (electromagnetic actuator or PCal, as shown in this paper, 
but also Newtonian calibrator as shown in~\cite{bib:2020_NCalO3}).
This bias is within $\pm3.9\%$ in modulus and $\pm30~\text{mrad}$ in phase,
the biggest bias being around 140~Hz for the modulus and around 90~Hz for the phase.
The estimation of this systematic bias has some uncertainties coming from the estimation of the hardware injection signal using the actuator models. 
They are estimated to a maximum of 1.84\% in modulus and 6.7~mrad in phase for the electromagnetic actuators 
and 1.73\% in modulus for the photon calibrators (see table~\ref{tab:MirUncertainties}).

In addition, some time variations of this systematic bias must be taken into account.
Weekly cadence measurements have shown variations at the level of 0.6\% in modulus and 6~mrad in phase (see table~\ref{tab:hrechinj_systO3}).
However, continuous monitoring of the bias at few frequencies in the Virgo sensitive band,
and in particular around the frequencies where the bias is the largest, 
have shown that the variations were a bit larger, of the order of 1.4\% and 14~mrad during O3a, and of 1.1\% and 11~mrad during O3b (see section~\ref{sec:hrecerrors_permmoni}).

The conservative frequency-independent symmetric uncertainties on the amplitude and on the phase of the reconstructed strain channel $h(t)$ during O3
have been estimated summing quadratically the maximum estimated bias (3.9\%, 30~mrad), 
its uncertainty coming from actuator model knowledge (1.8\%, 7~mrad)
and its time variations estimated continuously during O3 (1.4\%, 14~mrad). 
The total uncertainty on the $h(t)$ amplitude is thus 4.5\% and
the total uncertainty on the $h(t)$ phase is estimated to 34~mrad.\\

As described in section~\ref{sec:hrecerrors_weekly_broadband}, the uncertainties are larger around 50~Hz, especially during O3b.\\


An uncertainty on the absolute timing of the reconstructed Virgo strain channel $h(t)$ has been also estimated.
Comparing the reconstructed strain channel to the hardware injections (see section~\ref{sec:hrecerrors_weeklyO3a}), a timing bias estimation of 7~\mus\ has been found.
The uncertainty on the timing of the PCal hardware injections has been estimated to 3~\mus\ during O3a, 
with an additional error of 2.8~\mus\ during O3b (see section~\ref{sec:PCalCalibration}).
During O3b, an additional 1~\mus\ error must be added after the detector photodiodes were changed early 2020.
Summing quadratically these sources of errors, the total uncertainty on the timing of the reconstructed strain channel has been estimated at 8~\mus.\\

\lr{The uncertainties on the Virgo reconstructed strain channel $h(t)$ in the range 20~Hz to 2~kHz during the O3~run are conservatively given as frequency-independent uncertainties at the values summarized in table~\ref{tab:hrec_final_uncertainty}.
These uncertainties encompass the $h(t)$ frequency-dependent bias, the uncertainties on this bias and its variations as a function of time.
In this sense, they represent upper and lower bounds on the $h(t)$ uncertainty:
they overestimate the uncertainty if they are interpreted as 1-sigma deviations at each frequency.
}

\begin{table}[bt]
\centering
\begin{tabular}{|l|l|c|c|c|}
\cline{3-5} 
\multicolumn{2}{c|}{} & 20 Hz - 2 kHz  & 46 to 51 Hz & 49.5 to 50.5 Hz \\
\hline 
\multirow{3}[0]{*}{O3a}   & $h(t)$ amplitude & \multicolumn{2}{c|}{$\pm 5\%$}              & \multirow{3}[0]{*}{do not use} \\
                          & $h(t)$ phase     & \multicolumn{2}{c|}{$\pm 35\,\text{mrad}$}  &  \\
                          & $h(t)$ timing    & \multicolumn{2}{c|}{$\pm 10\,$\mus\ }       &  \\
\hline
\multirow{3}[0]{*}{O3b}   & $h(t)$ amplitude & $\pm 5\%$             & $\pm 40\%$             & \multirow{3}[0]{*}{do not use} \\
                          & $h(t)$ phase     & $\pm 35\,\text{mrad}$ & $\pm 600\,\text{mrad}$ & \\
                          & $h(t)$ timing    & $\pm 10\,$\mus\       & $\pm 10\,$\mus\        & \\
\hline
\end{tabular}
\caption{ \label{tab:hrec_final_uncertainty} 
Summary of the uncertainty on the reconstructed Virgo detector strain $h(t)$ during the run O3:
it includes both the $h(t)$ bias and the uncertainties on this bias.
For both periods O3a and O3b, the uncertainties are given for the $h(t)$ amplitude, phase and timing in the band 20~Hz to 2~kHz.
The given uncertainties are mainly frequency-independent, except for some bands close to 50~Hz where uncertainties are larger.
}
\end{table}

%% file: conclusion.tex
\section{Conclusion}

We have described the Advanced Virgo detector calibration,
the $h(t)$ reconstruction and the linear noise subtraction methods
used in the O3 observing run in 2019/2020. 
For the first time, the photon calibration technique was used as reference
for the Virgo detector calibration, 
after the Virgo photon calibrators have been cross-calibrated with LIGO ones. 
This technique was used to calibrate the NE, WE, BS, PR, NI and WI mirror and marionette actuators.
The actuator models are inputs for the reconstruction of the $h(t)$ detector strain signal.
The $h(t)$ bias and associated uncertainty have been estimated from
weekly measurements, as during~O2, but also from 
continuous hardware injections setup for~O3. 
Independent calibration methods, based on the old free Swinging Michelson technique
and on the new Newtonian calibration technique, have been used to cross-check the results.
For the O3 run, we estimated the upper limit on $h(t)$ bias and associated uncertainty 
to be 5\% in amplitude, 35~mrad in phase and 10~\mus\ in timing in most of the sensitive frequency band 20-2000~Hz,
with an exception around 50~Hz as detailed in section~\ref{sec:hrecerrors_conclusions}.
The $h(t)$ time series will be available publicly in GWOSC~\cite{bib:GWOSC}. 
The time series is called V1:Hrec\_hoft\_16384Hz for most of the data (online reconstruction),
and it is called V1:Hrec\_hoft\_V1O3ARepro1A\_16384Hz for the last two weeks of September~2019 for which the data were reprocessed to optimize noise subtraction.

Most of the transient gravitational wave events 
detected in the O3 observing run and described in~\cite{bib:GWTC2} 
had signal-to-noise ratio below~5 in the Virgo detector.
Their detection is insensitive to the few-percent level calibration uncertainties achieved~\cite{bib:2009_Lindblom_CalibAccuracy}. 
However, improvements in the astrophysical parameter estimation are on-going,
in particular to take into account the frequency-dependent calibration bias and uncertainties~\cite{bib:2021_Vitale_physiCal}.
It confirms the need for providing more accurate calibration and frequency-dependent estimation of the uncertainties \lr{for future observation periods}.

In view of the next observing rung, O4, to start end 2022, the PCal system is being improved to reduce its calibration uncertainties. 
However, the $h(t)$ bias and associated uncertainty are dominated by contributions from other sources in the different actuator calibration steps and in the $h(t)$ reconstruction. 
In addition, the Newtonian calibration technique is being further developed with the goal to provide another calibration method with similar uncertainties as those from the PCal for the O4 run, leading to a more precise understanding of the $h(t)$ bias and uncertainties.
With two such precise estimations of the $h(t)$ bias and uncertainties in the most sensitive band of the Virgo detector, frequency-dependent bias and uncertainties will be derived.

%% file: acknowledgments.tex
The authors gratefully acknowledge the Italian Istituto Nazionale di Fisica Nucleare (INFN),  
the French Centre National de la Recherche Scientifique (CNRS) and
the Netherlands Organization for Scientific Research (NWO), 
for the construction and operation of the Virgo detector
and the creation and support of the EGO consortium.
The authors also gratefully acknowledge research support from these agencies as well as by 
the Spanish  Agencia Estatal de Investigaci\'on, 
the Consellera d'Innovaci\'o, Universitats, Ci\`encia i Societat Digital de la Generalitat Valenciana and
the CERCA Programme Generalitat de Catalunya, Spain,
the National Science Centre of Poland and the European Union – European Regional Development Fund; Foundation for Polish Science (FNP),
the Hungarian Scientific Research Fund (OTKA),
the French Lyon Institute of Origins (LIO),
the Belgian Fonds de la Recherche Scientifique (FRS-FNRS), 
Actions de Recherche Concertées (ARC) and
Fonds Wetenschappelijk Onderzoek – Vlaanderen (FWO), Belgium,
the European Commission.
The authors gratefully acknowledge the support of the NSF, STFC, INFN, CNRS and Nikhef for provision of computational resources.
We would like to thank all of the essential workers who put their health at risk during the COVID-19 pandemic, without whom we would not have been able to complete this work.